\documentclass[]{mn2e}
\usepackage{psfig, epsf, epsfig}
\title[Formation of GC systems]{The origin of globular cluster systems from cosmological simulations}
\author[K. Bekki,  H. Yahagi, M. Nagashima,   and D. A. Forbes]
       {Kenji Bekki${}^1$\thanks{E-mail: bekki@bat.phys.unsw.edu.au},
        Hideki Yahagi${}^2$\thanks{E-mail: hideki.yahagi@nao.ac.jp},
        Masahiro Nagashima${}^3$\thanks{E-mail: masahiro@nagasaki-u.ac.jp},
        and Duncan A. Forbes${}^4$\thanks{E-mail: dforbes@astro.swin.edu.au} \\ 
        ${}^1$School of Physics, University of New South Wales,
              Sydney, NSW 2052,  Australia\\
        ${}^2$Division of Theoretical Astronomy, 
              National Astronomical Observatory,
              2-21-1 Osawa, Mitaka, Tokyo 181-8588, Japan\\
        ${}^3$Faculty of Education, Nagasaki University, Nagasaki,
              852-8521,  Japan \\
        ${}^4$Centre for Astrophysics \& Supercomputing,
Swinburne University of Technology,
Hawthorn, VIC 3122, Australia}

\begin{document}

\date{Accepted, Received 2005 May 13; in original form }

\pagerange{\pageref{firstpage}--\pageref{lastpage}} \pubyear{2005}

\maketitle

\label{firstpage}

\begin{abstract}

We investigate the structural, kinematical, and chemical properties
of globular cluster systems (GCSs) in galaxies of different Hubble types
in a self-consistent manner based on
high-resolution cosmological N-body simulations combined with
semi-analytic models  of galaxy and globular cluster (GC) formation.
We focus on correlations between  the 
physical properties of GCSs and those of their host galaxies 
for $\sim 10^5$ 
simulated galaxies located at the centres of dark matter halos (i.e. we do not 
consider satellite galaxies in sub-halos). 
Our principal results, which can be tested against observations,
are as follows.
The majority ($\sim$ 90\%) of GCs currently
in halos are formed in low-mass galaxies at  redshifts
greater than 3 
with mean formation redshifts of z = 5.7 (12.7 Gyrs ago) and 4.3
(12.3 Gyrs ago) for metal-poor GCs  (MPCs) 
and metal-rich GCs (MRCs) 
respectively.
About 52 \% of galaxies with GCs show clear bimodality in their 
metallicity distribution functions, 
though less luminous galaxies with 
$M_{\rm B}$ fainter than $-17$
are much less likely  to show bimodality owing to little or no MRCs.
The number fraction of MRCs does not depend on Hubble type
but is generally smaller for less luminous galaxies. 
The specific frequencies ($S_{\rm N}$) of GCSs  are typically
higher in ellipticals ($S_{\rm N} \sim 4.0$) 
than in spirals ($S_{\rm N} \sim 1.8$),
and higher again ($S_{\rm N} \sim 5.0$) for galaxies
located at the centers of clusters of galaxies.
The total number  of GCs per unit halo mass 
does not depend strongly on $M_{\rm B}$ or 
Hubble type of the host galaxy.
The mean metallicities of MPCs and MRCs
depend on $M_{\rm B}$ 
such that they are higher in more luminous galaxies,
though the dependence is significantly weaker for MPCs.
The spatial distributions of MRCs are more compact than those of MPCs and
we find that the half-number radii of MPCs ($r_{\rm e, mpc}$)
correlate  with the halo masses ($M_{\rm h}$) such that 
$r_{\rm e, mpc} \propto {M_{\rm h}}^{0.18}$.
There is no significant difference in velocity dispersions
between MPCs and MRCs.
We qualitatively compare  our results to
observational data where possible.
Finally, we discuss these results in the wider context of galaxy formation
and evolution.

\end{abstract}


\section{Introduction}

The physical properties of GCSs in galaxies
have long been considered to be ``fossil records''
that contain vital information on galaxy formation
and evolution (e.g., Searle \& Zinn 1978;
Harris 1991; Ashman \& Zepf 1998;
West et al. 2004).
For example, the observed bimodal colour
distributions, higher  specific frequencies ($S_{\rm N}$)
and higher fraction of metal-poor GCs (MPCs) 
in elliptical galaxies has motivated various 
formation scenarios for elliptical galaxies
(e.g., Ashman \& Zepf 1992; Forbes et al. 1997;
C\^ote et al. 1998).
Correlations between physical  properties of GCSs
and those of their host galaxies
also have provided  clues to the better understanding galaxy formation 
For example, 
Strader et al. (2004) showed that the 
mean colours of MPCs and MRCs 
correlate with the luminosity ($L$) of their host galaxy. 
From a survey of Virgo cluster early-type galaxies, 
Peng et al. (2006, P06) calculated from g--z colours, the relations
${\rm [Fe/H]}_{\rm gc}  \propto L^{0.16 \pm 0.04}$ for MPCs and 
${\rm [Fe/H]}_{\rm gc}  \propto L^{0.26 \pm 0.02}$ for MRCs.

In addition to the above chemical property of  GCSs,
the kinematical and structural properties of GCSs have been 
investigated both for MPCs and MRCs 
(Kissler-Patig \& Gebherdt 1998;
C\^ote et al. 2001;
Rhode \& Zepf 2004; Richtler et al. 2004
Peng et al. 2005;
Bergond et al. 2006; 
Bridges et al. 2006; 
Pierce et al. 2006;  
Romanowsky 2006;
Hwang et al. 2007;
Woodley et al. 2007).
Peng et al. (2005) showed that the GCS in NGC 5128 
has  a significant
amount of global rotation
whereas Richtler et al. (2004) did not find any significant
rotation in the GCS of NGC 1399,
which suggests a great diversity in the kinematics of GCSs in 
early-type galaxies.
The slopes of power-law density profiles of GCSs are also observed 
to be diverse and only loosely 
correlated with the luminosity of their host galaxy
(e.g., Harris 1986; Ashman \& Zepf 1998).

In comparison with this remarkable progress in observational
studies of GCSs (Brodie \& Strader 2006), 
theoretical modeling has not been as 
well developed as to  provide useful predictions
that can be compared with the wide range of  observations on GCS properties.
This is mainly because,
both {\it subpc-scale} formation processes
of GCs in galaxies and {\it kpc- and Mpc-scales}
merging/interaction processes over 
a Hubble time 
need to be modeled in a fully self-consistent manner
in order that the GCS properties and GCS-host relations can be
examined quantitatively.

Beasley et al. (2002; B02) first investigated the physical
properties of GCSs and their correlations with those of host
galaxies based on a semi-analytic model of hierarchical galaxy
formation in a $\Lambda$CDM universe. B02 however could not discuss
the structural and kinematical properties of GCSs owing to the
limitations of the adopted semi-analytic model. Although previous
collisionless N-body simulations with GCs have provided some
reasonable explanations for the origin of structural and
kinematical properties of GCSs (e.g., Bekki et al. 2005; Bekki \&
Forbes 2006), they did not model the chemical properties of GCSs.
Kravtsov \& Gnedin (2005) carried out a high resolution gas +
N-body simulation of GC formation in a $\Lambda$CDM universe. In
their model of a Milky Way like galaxy, GCs first formed at
redshift $\sim$ 12 with peak formation occurring at z $\sim$
4. However they only simulated MPCs and only from formation to a
redshift of 3.  Thus no simulations have yet addressed the
structural, kinematical, and chemical properties of GCSs in a
self-consistent cosmologically motivated simulation from
formation to the current epoch.

The purpose of this paper is  to investigate 
{\it  both the dynamical and chemical properties}  of GCSs
based on cosmological N-body simulations 
with semi-analytic models of galaxy formation
and thereby compare the results with
the above-mentioned  growing number of observational constraints on GCS-host
relations.
We focus particularly on the following physical properties
of GCSs in galaxies : (i) GC metallicity distribution functions (MDFs), 
(ii) number fractions of MRCs ($f_{\rm mrc}$),
(iii) number fractions  of GCS with bimodal MDFs ($f_{\rm bimo}$),
(iv) specific frequencies of GCs ($S_{\rm N}$),
(v) mean GCS metallicities (${\rm [Fe/H]}$),
(vi) GCS half-number radii ($r_{\rm e}$),
and (vii) GCS velocity dispersions ($\sigma$).
We mainly investigate correlations between these GCS properties
and their host properties such as $M_{\rm B}$ 
and Hubble type.
Here we do not discuss other important
observational results of GCSs, such
as the blue tilt (e.g., Strader et al. 2006)
or the GC luminosity function dependence on galaxy luminosities
(e.g., Jord\'an et al. 2006).
Some of these aspects have been already discussed in our previous
papers (e.g., Bekki et al. 2007a, B07). Here we also restrict our
analysis to central galaxies and do not consider satellites
in sub-halos.

The plan of this paper is as follows: In the next section 
we describe our numerical method of cosmological N-body simulations, 
semi-analytic model of galaxy formation,
and formation model for GCs.
In \S 3, we present our numerical results on  
GCS properties in galaxies with different luminosities and
Hubble types. 
In \S 4, we discuss our results of GCS-host relations  in the context of
galaxy formation and evolution.
We summarize our  conclusions in \S 5.

\begin{table*}
\centering
\begin{minipage}{185mm}
\caption{Model parameters in the simulation.}
\begin{tabular}{ccccccccccc}
$\Omega$
& $\Lambda$
& $H_{0}$  (km s$^{-1}$ Mpc$^{-1}$)
& ${\sigma}_{8}$
& {$M_{\rm T}$ 
\footnote{The total mass of a simulation
in units of $10^{14}$  ${\rm M_{\odot}}$.}}
&  { $R_{\rm T}$ 
\footnote{The box size of a simulation in units of
$h^{-1}$ Mpc }}
&  { $z_{\rm i}$
\footnote{The redshift ($z$) at which a simulation starts.}}
& {  $z_{\rm trun}$
\footnote{The redshift of truncation of GC formation:
GC formation is completely
truncated in halos that are  virialized later than the redshift.}}
& { $\alpha$
\footnote{The coefficient in the functional form describing
the dependence of GC formation rate on the mass ratios of merging two galaxies.}}
& { $\beta$
\footnote{The coefficient in the functional form describing
the dependence of GC formation rate on the gas mass fractions of galaxies.}}
& {SNe feedback 
\footnote{The details of the strong and weak feedback effects are given
in N05.}} \\
0.3 & 0.7 &  70  & 0.9  &  4.08 & 70 & 41 & 6 & 0.02 & 0.05  & strong  \\
\end{tabular}
\end{minipage}
\end{table*}

\section{The model}

We consider that GCs can be formed within  any galaxy at any redshift,
if the physical conditions required for GC formation are satisfied
in the galaxies. We therefore investigate GC formation rates in any
virialized halo 
in a high-resolution
cosmological simulation based on 
a $\Lambda$CDM cosmology model.
The physical properties of GCs (e.g., metallicities)
are determined by their host
galaxies at the epoch of their formation.
GCs formed in low-mass galaxies at high $z$ can be tidally
stripped during the hierarchical merging of galaxies
to finally become GCs within  a giant galaxy at $z=0$. 
Since numerical methods and techniques of the present  cosmological
simulations and those of semi-analytic models of galaxy formation
have been given in our previous papers 
(e.g., Yahagi et al. 2004; Nagashima et al. 2005, N05),
we only briefly describe them in the present study.
We here focus on (i) how to identify GCs in virialized dark matter halos
and (ii) how to allocate physical properties
to the identified GCs 
based on their host galaxy properties.

\subsection{Simulations}

We simulate the large scale structure of GCs
in a $\Lambda$CDM Universe with ${\Omega} =0.3$,
$\Lambda=0.7$, $H_{0}=70$ km $\rm s^{-1}$ ${\rm Mpc}^{-1}$,
and ${\sigma}_{8}=0.9$
by using the Adaptive Mesh Refinement $N-$body code developed
by Yahagi (2005) and Yahagi et al. (2004),
which is a vectorized and parallelized version
of the code described in Yahagi \& Yoshii (2001).
We use $512^3$ collisionless dark matter (DM) particles in a simulation
with the box size ($R_{\rm T}$) of $70h^{-1}$Mpc and the total mass
 ($M_{\rm T}$) of  $4.08 \times 10^{16} {\rm M}_{\odot}$.
We start simulations at $z_{\rm i} =41$ and follow it until $z=0$
in order to investigate the physical properties
of GCs outside and inside of virialized dark matter halos.
We use COSMICS (Cosmological Initial Conditions and
Microwave Anisotropy Codes), which is a package
of Fortran programs for generating Gaussian random initial
conditions for non-linear structure formation simulations
(Bertschinger 1995, 2001).

Our method of identifying  GCs (or ``GC particles'')
and following their evolution
is described as follows.
Firstly, we select virialized dark matter subhalos at a
given redshift
by using the friends-of-friends (FoF) algorithm (Davis et al. 1985)
with a fixed linking length of 0.2 times the mean DM particle separation.
The minimum particle number $N_{\rm min}$ for halos is set to be 10.
For each individual virialized subhalo,
the central
particle is labeled
as a ``GC'' particle.
This procedure for defining GC particles
is based on the assumption that energy dissipation via radiative cooling
allows baryons to fall into the deepest potential well of dark matter halos
and finally to be converted into GCs and stars.

The adopted initial distributions of GCs with respect to
their host halos  would be  oversimplified, given that possible
candidates of forming GCs are not necessarily in the central
regions of galaxies at $z=0$. 
We  think that the adopted assumption 
can be regarded as  reasonable,
because  previous observations suggested that 
a significant fraction (or even all) of the Galactic GCs
originate from nuclei of the Galactic building blocks
at high $z$
(e.g., Zinnecker et al. 1988; Freeman 1993).
We stress that the predicted spatial distributions
of GCSs in galaxies at $z=0$ might well  weakly depend on the adopted
initial distributions of GCs within halos,
though previous simulations suggested that distributions
of GCSs at $z=0$ do not depend strongly on the adopted range
of reasonable initial GC distributions 
(Yahagi \& Bekki 2005).

Secondly, we follow  GC particles
until $z=0$ and thereby
derive their locations $(x,y,z)$
and velocities $(v_{\rm x},v_{\rm y},v_{\rm z})$. 
We then identify virialized halos at $z=0$ with the FoF algorithm
and investigate whether each GC is within the virial radius
($r_{\rm vir}$) of a halo.
If GCs are found to be within a halo, the 
physical properties of the GCS
are investigated.
If a GC is not in any halo,
it is regarded as an intergalactic GC.
We don't discuss these intergalactic 
(i.e., intra-group and intra-cluster) GCs further in this
current paper (see Yahagi \& Bekki 2005; Bekki \& Yahagi 2006
for details of such GCs),
though a growing number of observations have revealed
physical properties of these intergalactic GCs
(e.g., Bassino et al. 2006; Jones et al. 2006).

Recent theoretical works have suggested that
heating and gas loss  resulting from reionization 
can severely suppress star formation in low-mass galaxies during
reionization
(e.g., Susa \& Umemura 2004).
It is therefore highly likely that GC formation is truncated
in low-mass galaxies that are virialized after the completion of reionization.
In order to include the effects of 
the suppression of GC formation 
via reionization on the final properties of the
simulated GCSs, 
we adopt the following somewhat idealized
assumption: If a galaxy  is virialized after the completion of
reionization ($z_{\rm reion}$), then
GC formation is totally suppressed in the galaxy.
Therefore, GC particles formed 
in galaxies  with $z_{\rm vir}  < z_{\rm reion}$
are not 
considered in the physical
properties of the simulated GCSs. 
We define the truncation epoch of GC formation as $z_{\rm trun}$
(rather than $z_{\rm reion}$ for convenience)
in the present study.
Recent quasar
absorption-line studies give a lower limit of 6.4 for 
$z_{\rm reion}$  (Fan et al. 2003). 
Guided by these observations, we investigate the
model with $z_{\rm trun}=6$.

Thus the  GCS of a galaxy 
is a collection of GCs that are formed within  low-mass galaxies
(i.e., galaxy building blocks)
embedded in massive dark matter halos virialized at high redshifts.
The physical properties 
of the GCS of a galaxy therefore depends
on star formation histories, chemical evolution, and merging
histories of the building blocks.
Chemical evolution and star formation histories of building blocks
for galaxies are derived from  the semi-analytic model  
(N05) 
which are based on the merging histories of DM halos
derived  from our N-body simulation.

\subsection{Semi-analytic model}

The physical properties of a newly formed 
GC in a virialized halo at any redshift
are determined by those of its host galaxy at that redshift.
N05 constructed the Numerical Galaxy Catalog ($\nu$GC)
based on a semi-analytic model  combined with high-resolution $N-$body
simulations.
In the present study, we adopt the same semi-analytic model as
that used by N05 which includes various physical
processes associated with galaxy formation,
such as galaxy merging, radiative gas cooling,
star formation,  supernovae feedback,  and extinction by internal
dust. Since the methods and techniques
are given in detail by
N05, we do not describe them in the present study.

The semi-analytic model by N05 can reproduce many observations reasonably well,
such as cold gas mass-to-stellar luminosity ratios of spiral galaxies,
faint galaxy number counts, cosmic star formation rates, 
the Tully-Fisher relation
for bright spiral galaxies,
luminosity functions of local galaxies,
and colour-magnitude relations for massive and dwarf elliptical
galaxies. The successes and limitations of this model in 
explaining observed galactic properties are given in  N05.

Here we present the results for the properties of GCs from our
semi-analytic model and compare them where possible to
observations.  We focus on results from the model with the
``strong feedback effects'' (N05) in which galaxies with smaller
rotational velocities are much more strongly influenced by
heating. An important caveat in this work is that only galaxies
located at the centers of halos are analyzed in the present
study, i.e. we do not investigate satellite galaxies. Thus we
effectively focus on brightest cluster, brightest group and
isolated galaxies in this current work. Observational studies are
often dominated by non-central galaxies (satellites in this
context) and this should be born in mind when we compare our
model predictions to the observations below.

\subsection{Formation efficiencies of GCs}

We convert the star formation rate (SFR) in a galaxy into
a GC formation rate (GCFR) based on the physical properties
of the host galaxy at a given redshift.
The vast majority 
of stars are observed to form
in star clusters (SCs)  embedded within GMCs (Lada \& Lada 2003).
Strongly bound SCs can evolve into GCs or old open clusters 
whereas weakly bound, low-mass  ones can be disintegrated into 
field stars of the host galaxy. 
The mass fraction of  new SCs that  evolve into GCs
relative to all new SCs is denoted  as $C_{\rm eff}$ 
and is assumed to be dependent on the physical properties of their host
galaxy. 
Thus we define the GCFR as:
\begin{equation}
{\rm GCFR} = C_{\rm eff} \times {\rm SFR}, 
\end{equation}
where SFR is an output from the 
semi-analytic model.

\begin{figure}
\psfig{file=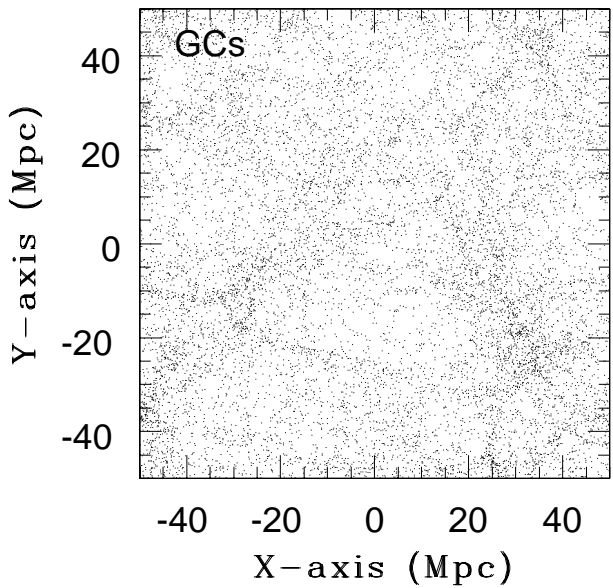,width=8.0cm}
\caption{
The distribution of GCs 
projected onto the $x$-$y$ plane  at $z=0$.
Here only GC particles that are at the very centers of GCSs
for  galaxies  
at $z=0$ are shown for convenience.
Thus the distribution describes
the large-scale distribution  of GCSs of galaxies  in the universe at $z=0$.
}
\label{Figure. 1}
\end{figure}

\begin{figure}
\psfig{file=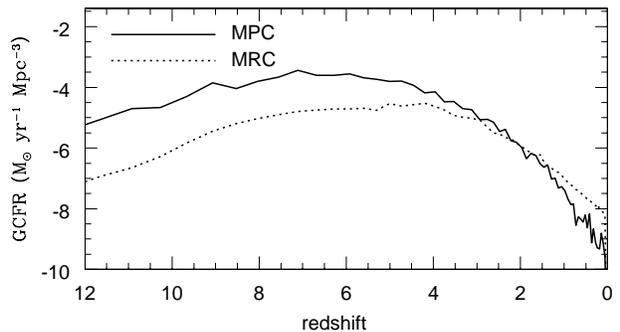,width=8.0cm}
\caption{
Cosmic evolution of GC formation rates (GCFRs) as a function of redshift ($z$)
for MPCs (solid) and MRCs (dotted).
The GCFRs are averaged over the volume of the simulation.
}
\label{Figure. 2}
\end{figure}

\begin{figure}
\psfig{file=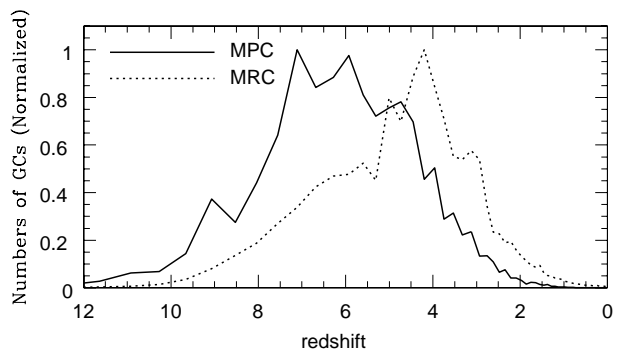,width=8.0cm}
\caption{
Total numbers of GCs formed at redshift $z$
for MPCs (solid) and MRCs (dotted). The numbers are normalized
by the maximum values
for $0 \le z \le 12$. 
}
\label{Figure. 3}
\end{figure}

We consider both observational results by Larsen \& Richtler (2000, LR00)
and simulation ones by Bekki et al. (2002, BFBC02) in order to determine
$C_{\rm eff}$  
in a physically reasonably way.
LR00 investigated correlations between
specific $U-$band cluster luminosities  $T_{\rm L}(U)$ 
of GCSs and their host
galaxy properties both for apparently isolated galaxies and
for interacting/merging ones.
LR00  found that (i) $T_{\rm L}(U)$ correlates with
SFR per unit area,  stellar surface brightness, and HI surface 
density,
and (ii) $T_{\rm L}(U)$ is more than an order of
magnitude higher in strongly starbursting mergers 
(e.g., NGC 1705) than apparently normal galaxies
that are forming young SCs (see Table 1 in LR00).
These results by LR00 imply that $C_{\rm eff}$ can be  higher 
in galaxies with higher surface densities and higher gas mass fractions
for a given galaxy luminosity.

Numerical simulations of GC formation in merging galaxies
(BFBC02)
found that $C_{\rm eff}$ depends on the mass ratios of the merging
spirals ($f_{\rm m}$ or $m_{2}$) and their gas mass fractions ($f_{\rm g}$) 
in such a way that  $C_{\rm eff}$ is higher
in mergers with larger $m_{2}$ and $f_{\rm g}$. 
These  results by  BFBC02 combined with LR00
imply that $C_{\rm eff}$ is a minimum for isolated galaxies with $m_2=0$
and maximum for equal mass  galaxy mergers with  $m_2=1$.
BFBC02 also found that $C_{\rm eff}$ is lower  in low surface brightness
galaxies for a given galaxy mass.
Galaxies with their halos virialized at higher $z$ 
have more compact disks 
and thus higher stellar densities for a given mass,
baryonic  mass fraction,  spin parameter, and halo circular velocity
in galaxy formation models based on
$\Lambda$CDM (e.g., Mo et al. 1998).
The above result by BFBC02 combined with that by Mo et al. (1998)
therefore strongly suggests that
 $C_{\rm eff}$ is likely to be higher for galaxies
formed at higher $z$. Thus $C_{\rm eff}$ is a function of the gas
fraction, the merger mass ratio and formation redshift.

We estimate $C_{\rm eff}$ as follows:
\begin{equation}
C_{\rm eff} = C_0 
F_{\rm g}(f_{\rm g}) F_{\rm m}(f_{\rm m})
F_{\rm z}(z). 
\end{equation}

We need to choose the forms of 
these three functions ($F_{\rm g}$, $F_{\rm m}$, and $F_{\rm z}$)
so that they are consistent with previous results by LR00 and BFBC02.
The normalization factor $C_0$ is determined as follows:
\begin{equation}
 C_0 = {F_{\rm g,max}}^{-1} {F_{\rm m,max}}^{-1} 
{F_{\rm z,max}}^{-1},
\end{equation}
where $F_{\rm g,max}$, $F_{\rm m,max}$, 
and $F_{\rm z,max}$
are maximum values of the above three functions. 
This method of normalization ensures that $C_{\rm eff}$ is always equal to
or less than 1.
If we adopt a value of $C_0$ significantly smaller than the one
given in the above equation,  
physical properties of the simulated GCSs (e.g., total numbers of GCs
in galaxies) can be much less consistent with observations.

In the present study, we choose elementary functions
for $C_{\rm eff}$.
Considering that (i) the dependence of $C_{\rm eff}$ on $f_{\rm m}$ 
appears to be non-linear (BFBC02) and (ii) 
$C_{\rm eff}$ is much higher in major mergers (LR02 and BFBC02),
we assume the following: 
\begin{equation}
F_{\rm m}(f_{\rm m})=1 - {(1+\alpha)}^{-1} + {(1+\alpha-f_{\rm m})}^{-1},
\end{equation}
where $\alpha$ is a parameter.
This formula ensures that $F_{\rm m}(f_{\rm m})$ has a minimum ($=1$)
at $f_{\rm m}=0$
corresponding to isolated disk galaxies 
and a maximum at $f_{\rm m}=1$ corresponding to major mergers.
Furthermore, if a small value of $\alpha$ ($<0.1$) is adopted,
$F_{\rm m}$ is more than 10, which corresponds to a strongly starbursting
major merger in LR00.
We adopt the same functional form for $F_{\rm g}(f_{\rm g})$:
\begin{equation}
F_{\rm g}(f_{\rm g})=1 - {(1+\beta)}^{-1} + {(1+\beta-f_{\rm g})}^{-1}.
\end{equation}
Thus,  $\alpha$ and $\beta$ are free parameters which
are chosen so that observations of GCSs can be self-consistently explained.

Mo et al. (1998) showed that the sizes ($R_{\rm d}$)  of disk galaxies
formed at redshift $z$ are inversely proportional to $H(z)/H_0$ for given disk and halo
properties (e.g., disk mass fraction and halo circular velocity).
Therefore  $R_{\rm d}$ can be  more compact for  disk galaxies
formed at higher redshifts so that mean stellar densities ${\Sigma}_{\rm s}$
within  $R_{\rm d}$ (${\Sigma}_{\rm s} \propto {R_{\rm d}}^{-2}$
for a given stellar disk mass)
are higher.
Considering the above dependences of disk properties on $H(z)/H_0$
(Mo et al. 1998),
we adopt the following functional form:
\begin{equation}
F_{\rm z}(z)={(H(z)/H_0)}^2 .
\end{equation}
We estimate $F_{\rm z}(z)$ for each halo at a given redshift
based on the virialization redshift ($z_{\rm vir}$) of the halo.

Here we adopt a simpler yet qualitatively reasonable
dependence  of $F_{\rm z}(z)$: we do not directly derive
the surface mass densities by assuming initial sizes of halos,
spin parameters,  and sizes of galaxies based on the 
results of the SAM by N05.
Although we can not discuss the importance of
these physical properties (e.g., initial spin parameters)
in GC formation in the present model,
we can show the importance of surface mass (or gas) densities
(i.e., the epoch of virialization) in GC formation of galaxies in a clearer
and more straightforward way thanks to the adopted model.
It should be stressed here that without introduction
of the $F_{\rm z}(z)$, most GCS properties can not be reproduced
well in the present model.

Although the adopted functional forms  of $F_{\rm m}$,
$F_{\rm g}$, and $F_z$ are  reasonable at least qualitatively,
no observations have been carried out which allow us to determine
whether the adopted functional form is quantitatively
consistent with observations.
Accordingly other functional forms
could be adopted so that observational properties of GCSs can be
reproduced s well.
Here we do not discuss the results of models
with different functional forms of $F_{\rm m}$,
$F_{\rm g}$, and $F_z$.

$T_{L}(U)$ in some merging galaxies
(e.g., NGC 1705) are observed to be as large as $15$,
which is a factor of $\sim20$ larger than
the average value ($=0.7$) of  $T_{L}(U)$ 
for non-merging galaxies with young GCs
(LR00). 
Also there is  a factor of $\sim 200$ difference
between the minimum (=0.07) and maximum (=15)
values of $T_{L}(U)$ for galaxies with young GC
candidates in LR00. 
Therefore $0.005 \le \alpha \le 0.05$ is a reasonable range
of $\alpha$ in $F_{\rm m}$.
We try to determine the most reasonable values of $\alpha$
and $\beta$
for which both 
the observed net formation efficiency of GCs (McLaughlin 1999)
and the number fraction of MRCs (Spitler et al. 2007)
can be well reproduced by the present simulation.

The net formation efficiency of GCs ($\epsilon$) in the present study 
is defined as follows: 
\begin{equation}
\epsilon=\frac{M_{\rm gc}}{M_{\rm star}},
\end{equation}
where $M_{\rm gc}$ and $M_{\rm star}$ are total masses of GCs
and stars, respectively,
which are formed in {\it all} building blocks in the simulation.
We define $f_{\rm mrc}$ as follows: 
\begin{equation}
f_{\rm mrc}=\frac{N_{\rm mrc}}{N_{\rm mpc}+N_{\rm mrc}},
\end{equation}
where $N_{\rm mpc}$ and $N_{\rm mrc}$ are numbers of MPCs and MRCs,
respectively.
Since we  find that the model with $\alpha=0.02$ and  $\beta=0.05$
can best reproduce the observations,
we discuss this model in this paper.

\begin{figure}
\psfig{file=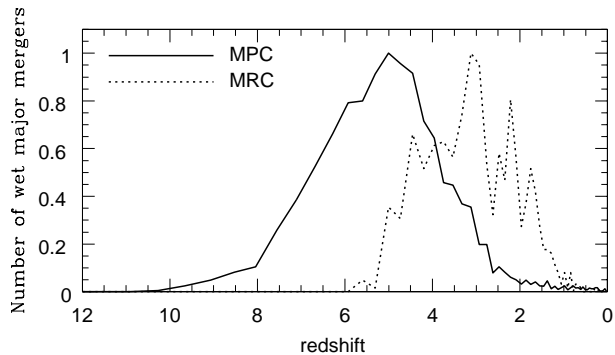,width=8.0cm}
\caption{
Number of gas-rich (``wet'')
major mergers that form MPCs (solid) and MRCs (dotted)
as a function of $z$.  Here wet major mergers are those with
$f_{\rm g} > 0.5$  and $f_{\rm m}>0.5$.
The numbers normalized by their maximum values for $ 0\le z \le 12$
are shown.
}
\label{Figure. 4}
\end{figure}

\begin{figure}
\psfig{file=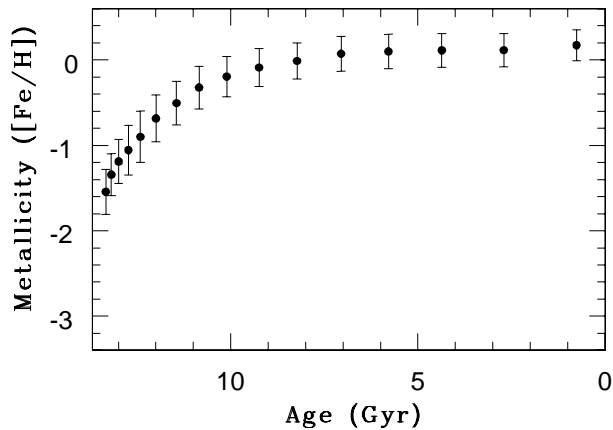,width=8.0cm}
\caption{
The age-metallicity relation for all GCs.
The mean metallicity of GCs in 
age bins are shown by  filled
circles. 
The $1 \sigma$ dispersion in the metallicities
of GCs  for each age bin is shown by an error bar.
}
\label{Figure. 5}
\end{figure}

\begin{figure}
\psfig{file=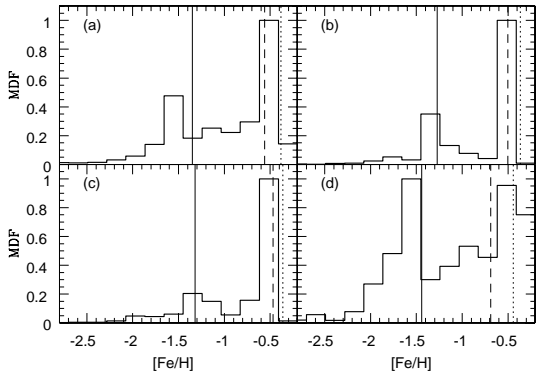,width=7.0cm}
\caption{
Metallicity distribution functions (MDFs) for four representative
massive galaxies with
${\log}_{10} ( \frac{ M_{\rm h} } { {\rm M}_{\odot} }) \ge 13.0$:
(a) $M_{\rm B}=-20.7$ mag and $B/T=1.0$,
(b) $M_{\rm B}=-21.3$ mag and $B/T=1.0$,
(c) $M_{\rm B}=-21.3$ mag and $B/T=1.0$,
and (d) $M_{\rm B}=-20.9$ mag and $B/T=1.0$.
The mean metallicities of MPCs, MRCs, and all GCs are shown
by solid, dotted, and dashed lines, respectively.
Although the shapes of MDFs are different in different galaxies,
they all clearly show bimodal MDFs.
}
\label{Figure. 6}
\end{figure}

\subsection{Main points of analysis}

The total number of virialized halos at $z=0$ is 95139,
among which only 12179 
(thus 12.8 \%) have GCs that are formed in their building
blocks virialized before $z=6$ (i.e., reionization).
The total number of {\it GC particles}  at $z=0$ 
is 998529 in the present simulation.
Fig. 1 shows the large-scale structure of GCs that
are in the very central regions of their host galaxies 
at $z=0$.  Although these central GCs in galaxies are  
relatively metal-rich with a mean metallicity of
${\rm [Fe/H] }=-0.18$ for the 12179 galaxies,
some fraction of them (37.5\%)  are metal-poor 
(${\rm [Fe/H] }<-1$).
Here we focus on correlations between physical properties
of GCSs and those of their host galaxies: we do not 
discuss the internal properties of individual GCSs.

In investigating GCS properties, 
we divide GCs into MPCs and MRCs according to their metallicities
([Fe/H]): those with ${\rm [Fe/H]} <-1$
and with ${\rm [Fe/H]} \ge -1$ are defined as MPCs
and MRCs, respectively.
We investigate the physical properties of GCSs separately
for these two GC subpopulations. 
We also discuss the bimodality in the metallicity distribution
functions (MDFs)
of  GCSs.
We show some examples of individual 
representative GCSs in Appendix A.

Following the morphological classification scheme by 
Simien \& de Vaucouleurs (1986),
we classify simulated galaxies into three different morphological 
types: E, S0, and Sp. 
We use  the simulated 
B-band bulge-to-disk luminosity ratio, 
B/D in the above morphological classification.
In this paper, 
galaxies with B/D $>$  1.52, 0.68 $<$  B/D $<$ 1.52, 
and B/D $<$ 0.68 are 
classified as elliptical (E), lenticular (S0), 
and spiral galaxies (Sp), respectively. 

\begin{figure}
\psfig{file=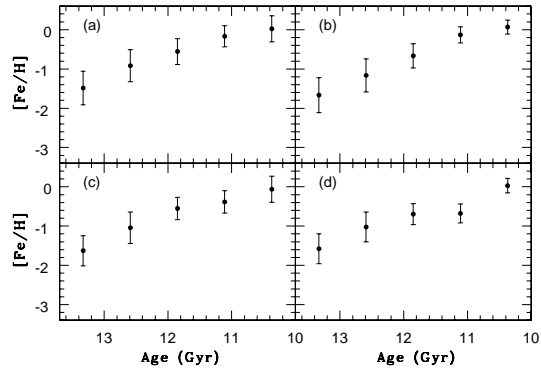,width=7.0cm}
\caption{
The age-metallicity relations (AMRs) of GCs for the four
galaxies shown in Fig. 6. 
The mean
metallicities
of GCs are shown for the five age bins.
In order to show the AMRs more clearly,
only GCs with ages larger than 10 Gyr are shown as few younger
GCs are present.
The $1 \sigma$ dispersion in the metallicities
of GCs  for each age bin is shown by an error bar.
}
\label{Figure.7}
\end{figure}

\begin{figure}
\psfig{file=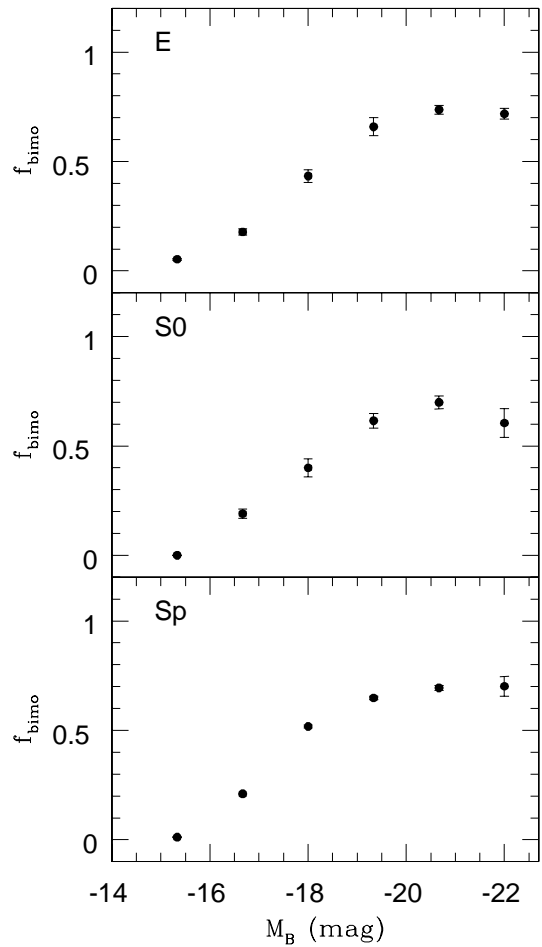,width=7.0cm}
\caption{
The number fractions of GCSs with bimodal MDFs ($f_{\rm bimo}$)
as a function of $M_{\rm B}$ for Es (top), S0s (middle), and
Sp (bottom).
The error bar in each bin is based on Poisson $\sqrt{N}$
statistics. 
}
\label{Figure.8}
\end{figure}


\begin{figure}
\psfig{file=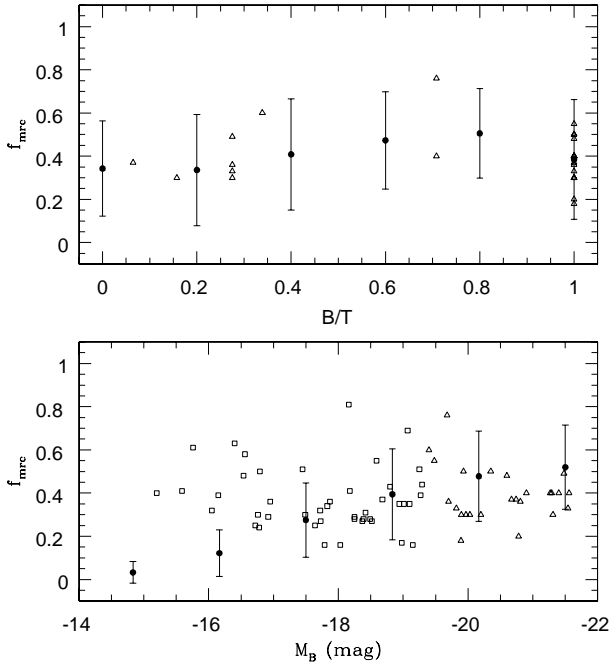,width=8.0cm}
\caption{
Distributions of the simulated GCSs in
the $f_{\rm mrc}$-$B/T$ plane (upper)
and the  $f_{\rm mrc}$-$M_{\rm B}$ one (lower),
where  $f_{\rm mrc}$ is the number fraction of MRCs in a GCS.
Triangles  and squares  represent the observational results
by Spitler et al. (2007) and by P06, respectively, whereas
circles  show the average values  of $f_{\rm mrc}$
in $B/T$ and $M_{\rm B}$ bins.
The $1\sigma$ dispersion
for each  bin is shown by an error bar.
}
\label{Figure. 9}
\end{figure}

\begin{figure}
\psfig{file=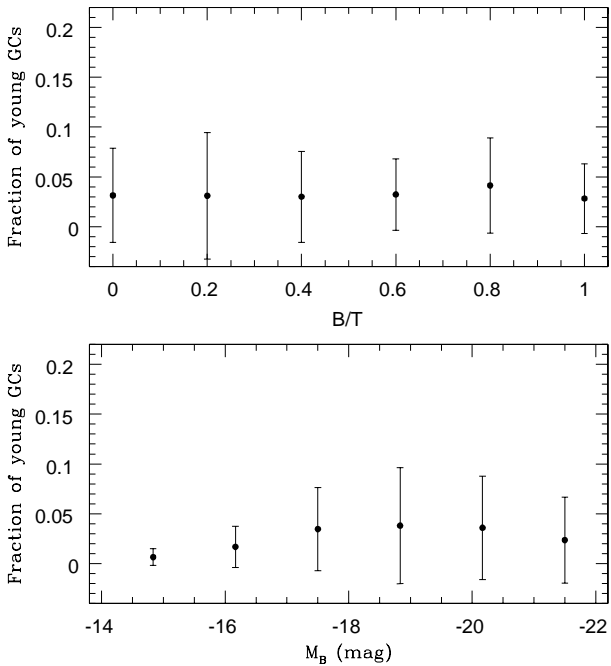,width=8.0cm}
\caption{
Distributions of the simulated GCSs on
the $f_{\rm y}$-$B/T$ plane (upper)
and on the  $f_{\rm y}$-$M_{\rm B}$ one (lower),
where $f_{\rm y}$ is the number fraction of young MRCs with ages
less than 8 Gyrs.
Circles show  the average values  of $f_{\rm y}$
in $B/T$ and $M_{\rm B}$ bins.
The $1\sigma$ dispersion
for each  bin is shown by an error bar.
}
\label{Figure. 10}
\end{figure}

\section{Results}

\subsection{Cosmic evolution}

Since our main focus is to discuss the results of correlations between
physical properties of GCSs and those of their host galaxies at $z=0$,
we do not discuss the 
cosmic evolution of GCS.
We do however  briefly summarize the cosmic evolution of GCS properties
{\it  averaged over all GCs formed in the simulation.}
Fig. 2 shows that the peaks of cosmic GCFRs are at $z \sim 7$ for MPCs
and at $z \sim 4$ for MRCs and that the GCFR is much higher
in MPCs than in MRCs at $z>2$, whereas it is higher in MRCs for $z<2$. 
Fig. 3 shows that number distributions of MPCs and MRCs normalized by
their maximum values have two different peaks,
which reflect the differences in cosmic evolution of GCFRs 
between  MPCs and MRCs.

Fig. 4 shows that  both MPCs and MRCs 
can be formed from gas-rich ($f_{\rm g}>0.5$),
major ($f_{\rm m}>0.5$) mergers for a wide range of redshifts
and that formation rates of MPCs and MRCs from wet major mergers 
peak at $z \sim 5$ for MPCs and  at $z \sim 3$ for MRCs.
As shown in Fig. 5,
the mean metallicities for all GCs (including both MPCs and MRCs)
steeply increase for ages $ > 10$ Gyr 
as a result of rapid chemical enrichment in the building blocks
of galaxies. 
The time evolution of the mean metallicities 
appears to be much less dramatic for ages $ < 10$ Gyr
owing to the slower chemical enrichment processes in galaxies. 
The dispersions in the mean metallicities 
($\sigma ({\rm Fe/H})$) are smaller 
for larger ages i.e., older GCs.  
We find $\sigma ({\rm Fe/H}) \sim 1.6$ dex for $ \sim 13$
Gyr and $\sigma ({\rm Fe/H}) \sim 0.1$ dex for $ \sim 0.8$ Gyr.

It is interesting to derive 
the age-metallicity relation  of GCs with ages $t$  (Gyr) older than
10 Gyr in the present simulation at $z=0$
in order to compare the simulated relation with 
the observed one when sufficient observational results are available.
The $\chi$-square fit to the simulation data gives
the following:
\begin{equation}
{\rm [Fe/H]} = 10.86-10.83 {\log}_{10}  t
\end{equation}
The model predicts the presence of young ($<3$ Gyr) GCs
with ${\rm [Fe/H]} > 0$, 
though the number fraction of such GCs 
is quite small. 
Most of these metal-rich GCs are found to be located in the central
regions of galaxies in the simulation.
It should be however stressed
that very old ($>13$ Gyr), metal-poor (${\rm [Fe/H]} < -1.6$) 
GCs can also be found in the very central regions of galaxies at $z=0$.

\begin{figure}
\psfig{file=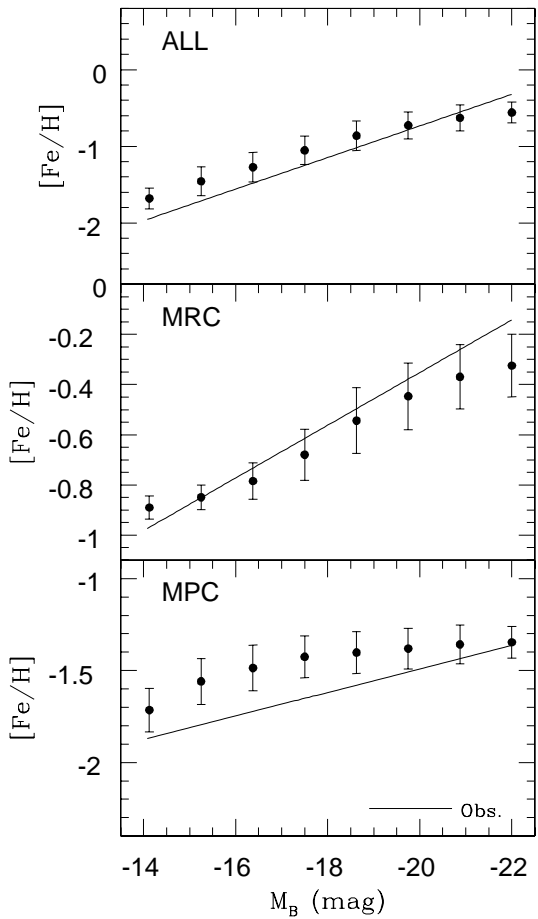,width=7.0cm}
\caption{
Distributions of the simulated GCSs in
the $M_{\rm B}$-${\rm [Fe/H]}$ plane for all GCs (top),
MRCs (middle), and MPCs (bottom).
Here ${\rm [Fe/H]}$ represents the mean metallicity of a GCS in
a galaxy.
Circles show the average values  of  ${\rm [Fe/H]}$
in a $M_{\rm B}$ bin.
The $1\sigma$ dispersion
for each  bin is shown by an error bar.
For comparison, the observed $M_{\rm B}$-${\rm [Fe/H]}$  relations
of P06 are shown by solid lines in the three panels.
}
\label{Figure. 11}
\end{figure}

\begin{figure}
\psfig{file=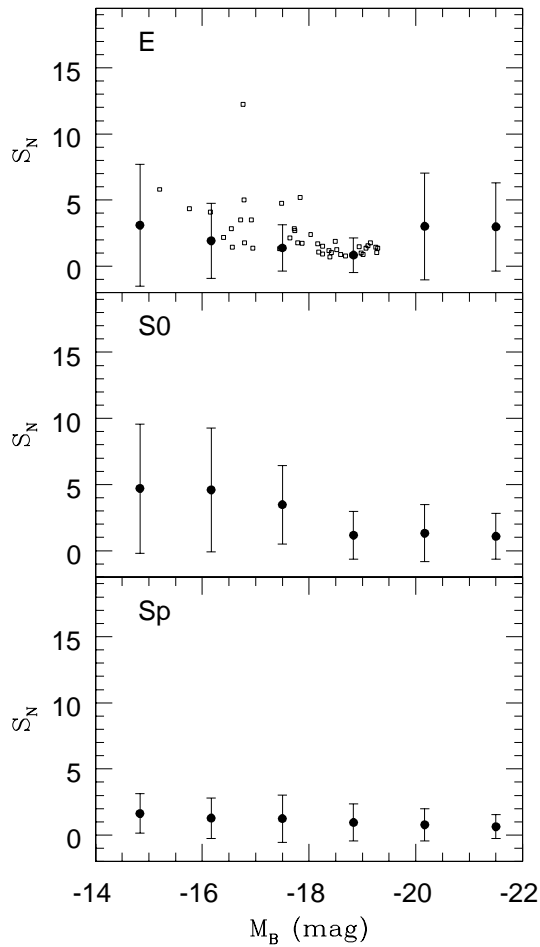,width=7.0cm}
\caption{
Distributions of the simulated GCSs in
the $S_{\rm N}$-$M_{\rm B}$ plane for E (top),
S0 (middle), and Sp (bottom).
Circles show the average values  of $S_{\rm N}$
in a $M_{\rm B}$ bin.
Open squares represent the observational results by P06,
for which we assume that $B-V=0.9$ for all Es to estimate
$S_{\rm N}$.
The $1\sigma$ dispersion
for each  bin is shown by an error bar.
}
\label{Figure. 12}
\end{figure}

\begin{figure}
\psfig{file=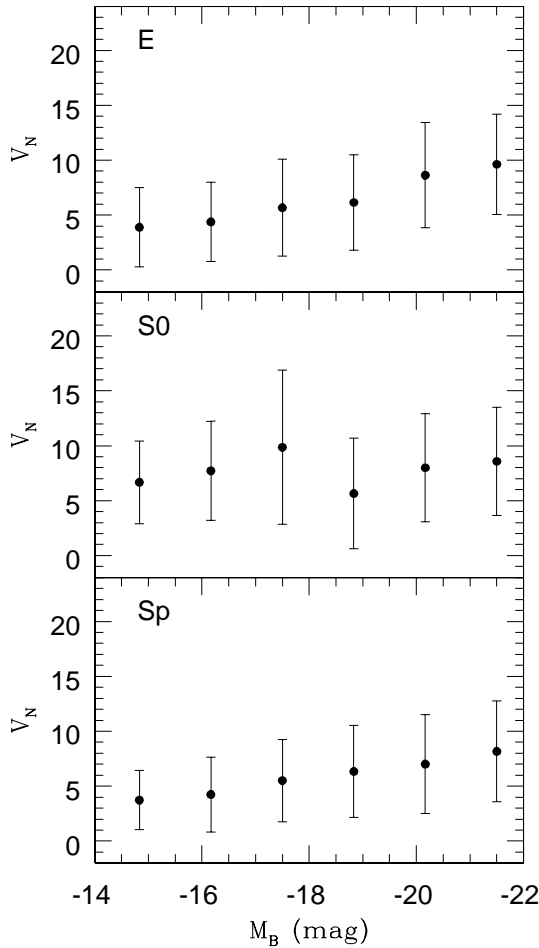,width=7.0cm}
\caption{
The same as Fig. 12 but for the  $V_{\rm N}$-$M_{\rm B}$ plane,
where  $V_{\rm N}$ is the number of GCs per halo mass.
}
\label{Figure. 13}
\end{figure}

\begin{figure}
\psfig{file=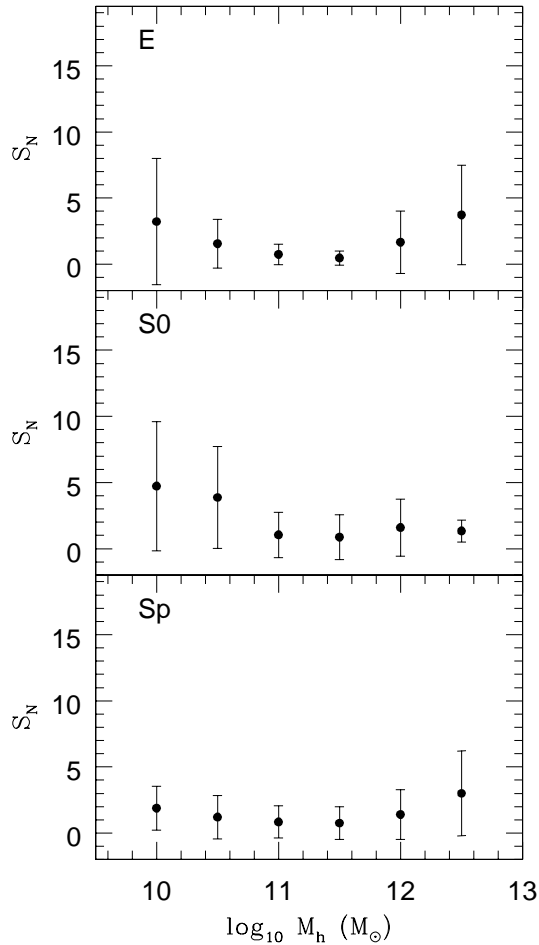,width=7.0cm}
\caption{
Distributions of the simulated GCSs in
the $S_{\rm N}$-$M_{\rm h}$ plane for E (top),
S0 (middle), and Sp (bottom).
Circles with error bars  show the average values  of $S_{\rm N}$
in $M_{\rm h}$ bins.
The $1\sigma$ dispersion
for each  bin is shown by an error bar.
}
\label{Figure. 14}
\end{figure}

\begin{figure}
\psfig{file=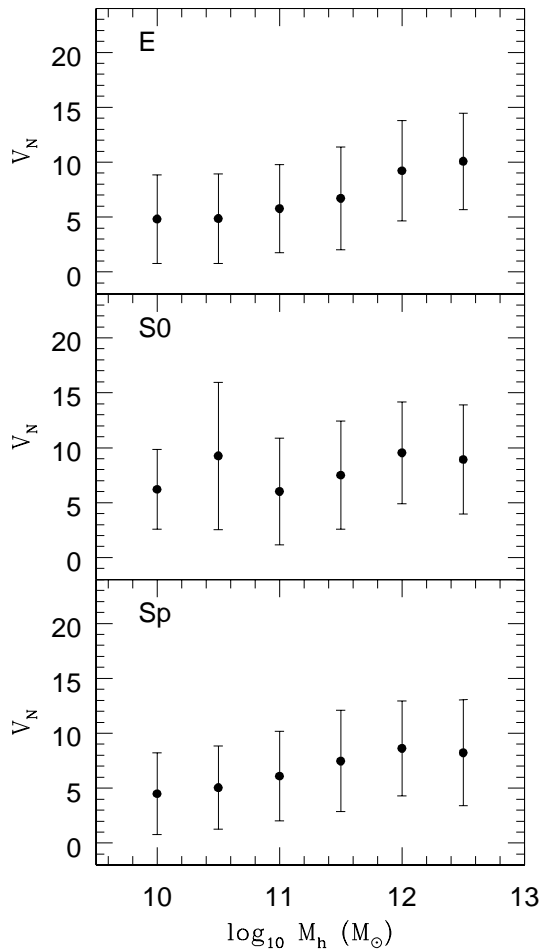,width=7.0cm}
\caption{
The same as Fig. 14  but for the  $V_{\rm N}$-$M_{\rm h}$ plane.
}
\label{Figure. 15}
\end{figure}

\begin{table}
\centering
\caption{Metallicity-luminosity relations. The observed (o; Peng
et al. 2006) and 
simulated (s) coefficients are given for MPC, MRC and all GCs,
where [Fe/H] = a + b M$_B$.}
\begin{tabular}{ccccc}
& ${\rm a}_{\rm o}$  
& ${\rm b}_{\rm o}$  
& ${\rm a}_{\rm s}$  
& ${\rm b}_{\rm s}$  \\
MPC & -2.771 & -0.064  &  -2.414 & -0.052  \\
MRC & -2.452 & -0.105  &  -2.056 & -0.081  \\
All & -4.854 & -0.206  &  -3.910 & -0.160  \\
\end{tabular}
\end{table}

\subsection{Metallicities and ages}

Fig. 6 shows that the GCSs in four early-type galaxies with $B/T
\approx 1$ at $z=0$ have bimodal MDFs with peaks below and
above ${\rm [Fe/H]} =-1$.  The locations of the two peaks and the
number ratios of GCs in each peak are different between
different galaxies, which implies that merging and star formation
histories of their building blocks are quite diverse in the four
galaxies.  The peak metallicities of GCSs reflect the gaseous
metallicities of the building blocks at the epoch when strong
starbursts were triggered mostly by gas-rich (``wet''), major
merging at high $z$ ($>3$).  Although these four
galaxies located in the centers of massive clusters of galaxies
with $M_{\rm h} > 10^{13} {\rm M}_{\odot}$ clearly show bimodal
MDFs, less luminous galaxies in the field and groups of
galaxies do not necessarily show bimodal MDFs in their GCSs.  The
diverse shapes of MDFs of these galaxies (including those
that show unimodal MDFs) are briefly discussed in Appendix A.

Fig. 7 clearly shows that younger GCs are likely to
be more metal-rich for all four galaxies
and that the age-metallicity relations are similar to one another.
Since  the colours of GCs depend on both age and metallicity,
these results suggest that the 
colour distributions of the simulated GCSs will be 
somewhat different from their MDFs. On the other hand, the GCSs
are dominated by very old GCs in each case. 
We plan to investigate whether the colour 
distributions of GCSs are consistent with observations
by combining the present GC formation model 
with a stellar population synthesis code and the
effects of photometric uncertainty in a future paper.

To investigate the number fractions ($f_{\rm bimo}$) of galaxies 
that reveal clear bimodal
MDFs 
we determine $f_{\rm bimo}$ 
as follows:
\begin{equation}
f_{\rm bimo}=\frac{ N_{\rm bimo} }{  N_{\rm gal}   },
\end{equation}
where $ N_{\rm bimo}$ and $  N_{\rm gal} $ are
the number of galaxies with obviously  bimodal GCS MDFs and
that of all galaxies, respectively.
In the present study,  an MDF with obvious bimodality
is  defined as revealing a clear metallicity peak at both low (${\rm [Fe/H]} <-1$  for MPCs)
and high metallicity (${\rm [Fe/H]}  \ge -1$ for MRCs).

Fig. 8 shows that more than 70\% of galaxies with $M_{\rm B}<-19$ mag
have bimodal MDFs in their GCSs
regardless of their Hubble types.
The number fraction of GCSs with bimodal MDFs ($f_{\rm bimo}$)
depends on $M_{\rm B}$ such that
less luminous galaxies have smaller $f_{\rm bimo}$
(i.e., less likely to have bimodal MDFs).
Considering that we do not 
include GCSs with no MPCs in estimating $f_{\rm bimo}$,
the above result means that
only MPCs are formed in 
the building blocks of these less luminous galaxies
with no bimodal MDFs. Furthermore gas-rich, major merging 
between high-density building blocks
does  not occur after their
gaseous metallicities becomes higher than ${\rm [Fe/H]} \sim -1$
in the formation histories of these galaxies.

 Fig. 9  shows that the number fraction of MRCs ($f_{\rm mrc}$)
does  not   depend on $B/T$.
Some very late-type spirals with $B/T<0.05$ have high values of $f_{\rm mrc}$
($>0.6$) whereas some ellipticals with $B/T \approx 1$ have low values
of $f_{\rm mrc}$ ($<0.1$), which suggests that
the origin of the Hubble types of galaxies  are not 
closely associated with the metal-rich fraction 
($f_{\rm mrc}$) in their GCSs.
 However $f_{\rm mrc}$ does depend on  $M_{\rm B}$ such that
 $f_{\rm mrc}$ is higher in more luminous galaxies,
though the dispersions in  $f_{\rm mrc}$ are quite large over
a wide range of $M_{\rm B}$.
Galaxies with $M_{\rm B}<-20$ mag cover a similar range for
observed galaxies in the $M_{\rm B}-f_{\rm mrc}$ plane (Spitler
et al. 2007).  The mean value of $f_{\rm mrc}$ for all GCSs in
the present simulation is 0.36, which is consistent with that
found by Spitler et al. (2007). We note that the Spitler et
al. sample includes non-central galaxies.

The mean formation epochs of MPCs and MRCs
for all GCs formed in the simulation between $z=41$ and $z=0$
are estimated to be 5.7 and 4.3,
respectively.
These results imply typical ages of MPCs and MRCs
at $z=0$ are 12.7 and 12.3 Gyr, respectively. This is consistent 
with the current spectroscopic measurements of extragalactic GCs
(e.g. Brodie \& Strader 2006). 
The mean
values of the  number fractions of young GCs
with ages less than 8 Gyrs ($f_{\rm y}$)
is very small ($\le$4\%) in the present simulation.
Fig. 10  shows that $f_{\rm y}$ does not depend on $B/T$.
The dependence of  $f_{\rm y}$ on  $M_{\rm B}$ 
appears to show no/little  trend: 
although  $f_{\rm y}$ appear to peak around 
$M_{\rm B} = -19$,
this apparent trend is of little 
statistical significance.
owing to the large dispersions.
Dispersions in the locations of galaxies
in the $B/T-f_{\rm y}$ and the $M_{\rm B}-f_{\rm y}$ planes
reflect the diversity in the formation epochs of young MRCs,
which are mostly via gas-rich major merging.

Next we investigate correlations between mean metallicities of GCs (simply
referred to as [Fe/H]) and $M_{\rm B}$ of their host galaxies.
These correlations are called ``metallicity-luminosity relations''
for convenience in the present study.
The observed metallicity-luminosity relations by P06, after 
transformation from observed colours, 
can be described as follows:
\begin{equation}
{\rm [Fe/H]}=a + b {M_{\rm B}}
\end{equation}
The values of the  observed coefficients $a_{\rm o}$ and $b_{\rm o}$
for MPCs, MRCs, and all GCs are summarized in Table 2.
The values of our simulated $a_{\rm s}$ and $b_{\rm s}$ 
using the same relation are listed also in Table 2
for comparison.

Fig. 11 shows that the mean metallicities of {\it all GCs} correlate well with
$M_{\rm B}$ such that the metallicities are higher in more luminous
galaxies. The simulated slope of the
${\rm [Fe/H]}-{\rm M}_{\rm B}$ correlation for all GCs
is consistent with the observed one by P06 for non-central
galaxies.
The simulated  ${\rm [Fe/H]}-{\rm M}_{\rm B}$ correlation
for MRCs is also consistent with the observed one
which suggests that the present formation model for MRCs
is realistic.
The  ${\rm [Fe/H]}-{\rm M}_{\rm B}$ correlation for
MPCs is, however, not so consistent with the observed one:
the slope of the correlation is too flat and the mean metallicities
of MPCs are systematically higher  than the observed ones for a wide
range of luminosities. 
One way to solve this inconsistency is discussed
later in \S 4.3.


\begin{table}
\begin{minipage}{85mm}
\centering
\caption{Mean $S_{\rm N}$ and $V_{\rm N}$ in different Hubble types}
\begin{tabular}{cccc}
&  Sp
&  S0 
&  E \\ 
$<S_{\rm N}>$ & 1.8 & 2.0  &  4.0 \\
$<V_{\rm N}>$ & 8.3 & 11.4  &  11.5  \\
\end{tabular}
\end{minipage}
\end{table}

\subsection{Specific Frequencies $S_{\rm N}$ and $V_{\rm N}$}

We investigate 
specific frequencies ($S_{\rm N}$) of GCSs in galaxies
with different Hubble types.
$S_{\rm N}$ is  defined as follows (Harris \& van den Bergh 1981):
\begin{equation}
S_{\rm N}=N_{\rm gc} 10^{0.4(M_{\rm v}+15)},
\end{equation}
where $N_{\rm gc}$ and $M_{\rm v}$ are the total number of globular clusters
in a galaxy and the $V-$band absolute magnitude of the galaxy, respectively.
We also investigate the number of GC per unit halo mass $V_{\rm N}$, 
which, following Spitler et al. (2007),  is defined as follows: 
\begin{equation}
V_{\rm N}=N_{\rm gc} {(\frac{M_{\rm h}}{10^{11} {\rm M}_{\odot} })}^{-1}
\end{equation}
where $M_{\rm h}$  is the total mass 
in a galaxy halo including dark matter.
The mean values of $S_{\rm N}$ and $V_{\rm N}$ 
in GCSs of  galaxies of different Hubble types
are summarized in Table 3. 
We estimate the $1\sigma$ dispersion in $S_{\rm N}$ and $V_{\rm N}$
for $M_{\rm B}$ and $M_{\rm h}$ bins to investigate the statistical
significance in the simulated correlations of
$S_{\rm N}$ and  $V_{\rm N}$ 
with  $M_{\rm B}$ and $M_{\rm h}$.
In estimating these dispersions,
we use the simulated galaxies with $S_{\rm N}$ ($V_{\rm N}$)
less than 20 to avoid  unreasonably large dispersions
caused by a very small number of galaxies with unusually
large $S_{\rm N}$ ($>50$).

Fig.  12 shows that $S_{\rm N}$ of E/S0s are typically higher
than those of spirals (Sp) for a given $M_{\rm B}$.  
Although more luminous Es with  $M_{\rm B}<-20$
have higher  $S_{\rm N}$ than
intermediate-luminosity Es  with   $-19$  $<M_{\rm B}<$ $-17$,
there is no such a  trend in spirals (Sp).
Faint galaxies with  $M_{\rm B}>-15$  show high  $S_{\rm N}$  ($>4$)
which for E/S0 is due to high halo mass-to-light-ratios
($M_{\rm h}/L_{\rm B}$).
The mean $S_{\rm N}$ for Sp, S0, and E populations are
1.8, 2.0,  and 4.0, respectively.

Fig. 13 shows that the dependences of  $V_{\rm N}$
on  the Hubble types of galaxies are not as strong
as those of $S_{\rm N}$: only a factor of $2-3$ difference
in  $V_{\rm N}$  between E/S0
galaxies and spirals for a given luminosity.
For  a given Hubble type,
$V_{\rm N}$ does not depend strongly on  $M_{\rm B}$.
These results mean that the 
numbers of GCs per unit halo mass ($M_{\rm h}$) does not
depend strongly on luminosity or the Hubble type
of their host galaxies.
The weak dependence of $V_{\rm N}$ on $M_{\rm h}$ was already
pointed by Bekki et al. (2006, B06), 
though $V_{\rm N}$ is estimated only for MPCs in B06.

 There is a weak tendency for more luminous Es with  $M_{\rm B}<-20$ mag
to have higher  $V_{\rm N}$,
though the dispersion in $V_{\rm N}$ is large.
S0s have $V_{\rm N}$ values $>8$ that are significantly
higher than those of spirals for a wide range of $M_{B}$,
suggesting that the higher  $V_{\rm N}$ in S0s
is not due to the truncation of star formation
in spirals as  $V_{\rm N}$ does not
change when a disk fades via truncation of star formation.
These results imply that if the observed  $V_{\rm N}$ are typically
higher in S0s than in spirals,
then only a minor fraction of spirals
can be transformed into S0s via truncation of star formation
in spirals.

\begin{figure}
\psfig{file=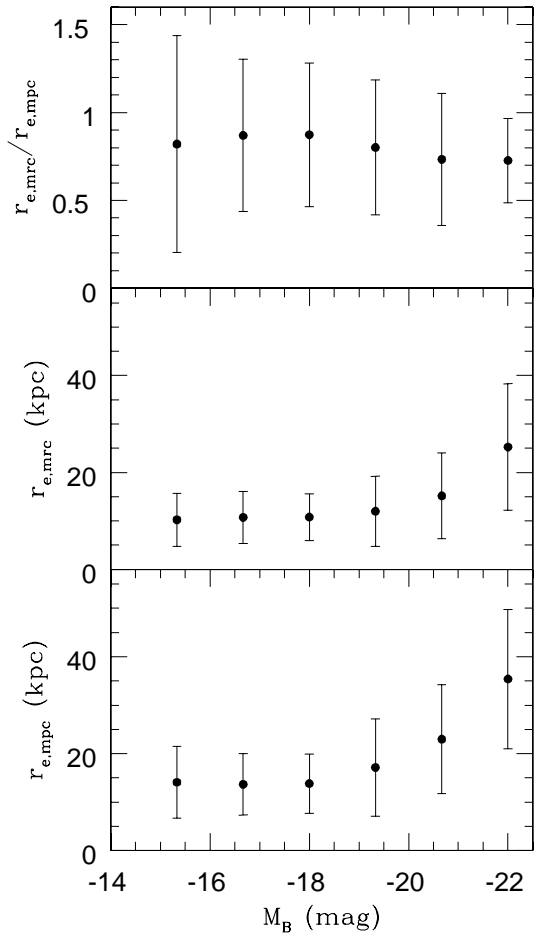,width=7.0cm}
\caption{
Dependences of 
$r_{\rm e, mrc}/r_{\rm e, mpc}$ (top) 
$r_{\rm e, mrc}$ (middle),  and
$r_{\rm e, mpc}$ (bottom)  
with $M_{\rm B}$.
The values averaged for GCSs in 
$M_{\rm B}$ bins  are shown by filled circles. 
The $1\sigma$ dispersion
for each  bin is shown by an error bar.
}
\label{Figure. 16}
\end{figure}

\begin{figure}
\psfig{file=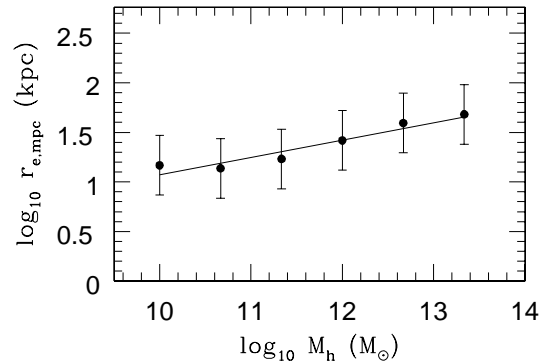,width=7.0cm}
\caption{
Dependence of $r_{\rm e, mpc}$ (i.e., the half-number radius of a GCS)
on $M_{\rm h}$ (halo mass of the host). 
The values averaged for GCSs in $M_{\rm h}$ bins are shown by
filled circles.
The $1\sigma$ dispersion
for each  bin is shown by an error bar.
The $\chi$-square fit to the simulation data is shown by a solid line.
}
\label{Figure. 17}
\end{figure}

\begin{figure}
\psfig{file=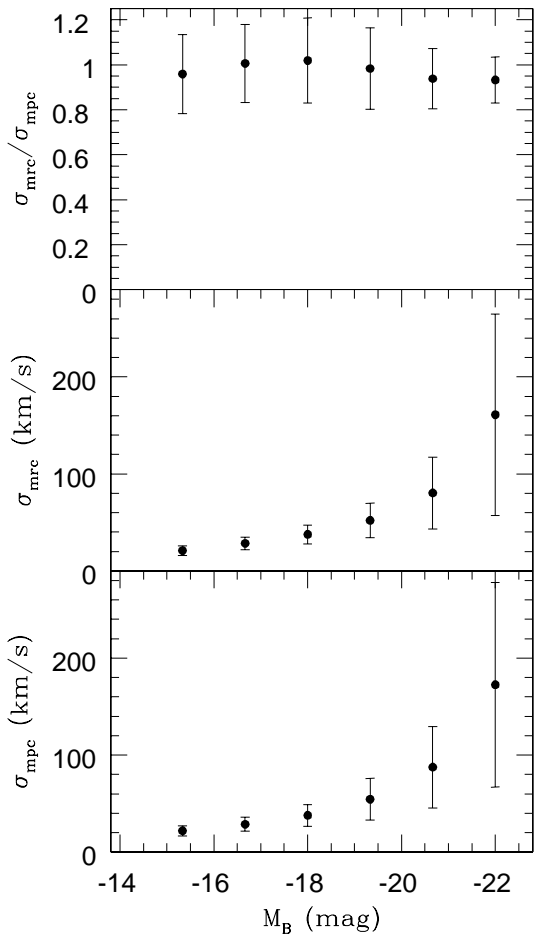,width=7.0cm}
\caption{
Dependences of 
${\sigma}_{\rm  mrc}/{\sigma}_{\rm mpc}$ (top) 
${\sigma}_{\rm  mrc}$ (middle),  and
${\sigma}_{\rm mpc}$ (bottom)  
on $M_{\rm B}$ in the simulation.
The values averaged for GCSs at each 
of the six $M_{\rm B}$ bins  are shown by triangles.
The $1\sigma$ dispersion
for each  bin is shown by an error bar.}
\label{Figure. 18}
\end{figure}

\begin{table*}
\centering
\begin{minipage}{185mm}
\caption{Mean properties of the simulated GCSs.}
\begin{tabular}{cccccccccc}
{$<z_{\rm f,mpc}>$
\footnote{The mean formation epoch  of all MPCs in the simulation.}}
& { $<z_{\rm f, mrc}>$ 
\footnote{The mean  formation epoch
of all MRCs in the simulation.}}
& { $<f_{\rm bimo}>$
\footnote{The mean number fraction of galaxies with GCSs showing 
clear bimodal MDFs.}}
& { $<f_{\rm mrc}>$
\footnote{The mean number fraction of MRCs in GCSs.}}
& { $<f_{\rm y}>$
\footnote{The mean number fraction of young GCs with ages
younger than 8 Gyrs in GCSs.}}
& {  $<S_{\rm N}>$ 
\footnote{The mean  $S_{\rm N}$ (GC number per unit luminosity)
in galaxies with GCs.}}
& {  $<V_{\rm N}>$ 
\footnote{The mean  $V_{\rm N}$ (GC number per unit halo mass)
in galaxies with GCs.}}
& {  $<s_{\rm eff}>$
\footnote{The mean ratio of half-number  radii of MRCs
to those of MPCs in GCSs.}}
& {  $<s_{\rm dis}>$
\footnote{The mean ratio of velocity dispersions
of MRCs to those of MPCs in GCSs.}}
& { $<{\epsilon}>$
\footnote{The mean net formation efficiency of GCs 
in galaxies.}} 
\\
5.7 & 4.3 &  0.52  & 0.36  &  0.04 & 2.11 & 11.25 & 0.84  & 0.98 & 0.0013   \\
\end{tabular}
\end{minipage}
\end{table*}

Fig. 14  shows an ``U-shape'' distribution in the $S_{\rm N}-M_{\rm h}$
results for Es, although
error bars are quite large (up to $\sim 5$ for low-mass halos).
This U-shape distribution
means that $S_{\rm N}$ is higher in smaller $M_{\rm h}$
below the threshold halo mass ($M_{\rm h, th}$)
of ${\log}_{10} M_{\rm h,th}=10^{11} {\rm M}_{\odot}$
whereas it is higher in larger  $M_{\rm h}$ above
$M_{\rm h, th}$. 
This simulated distribution for Es appears to be  broadly consistent with
the latest observation by Peng et al. (2008) for GCSs
in the Virgo cluster of galaxies.
This U-shape distribution can be also  seen in spirals,
though it is not so clear in S0s.
The $S_{\rm N}-M_{\rm h}$ relation for halo masses smaller than $M_{\rm h, th}$
is due mainly  to 
the fact that $M_{\rm h}/L_{\rm B}$ depends strongly on $M_{\rm h}$
(or on  $M_{\rm B}$).
The $S_{\rm N}-M_{\rm h}$ relation for halo masses larger  than $M_{\rm h, th}$
is due mainly to the fact
that numbers of GCs per unit halo mass
are higher in more massive galaxies.

As shown in Fig 15,
the U-shape distributions are less well defined for
the $V_{\rm N}-M_{\rm h}$ results 
of galaxies with different Hubble types.
However,  $V_{\rm N}$ in galaxies with  $M_{\rm h} > M_{\rm h, th}$
are significantly higher in Es, which means higher numbers of GCs per
unit halo mass in more massive Es.
More luminous spirals show higher  $V_{\rm N}$ for
 ${\log} M_{\rm h} < 10^{11.5} {\rm M}_{\odot}$.
It is not so clear why S0s have  $V_{\rm N}$ values 
higher than those of Es and Sps at low masses
${\log} M_{\rm h} < 10^{10.5} {\rm M}_{\odot}$.

The results in Figs. 14 and 15  thus show that
 $S_{\rm N}$ and  $V_{\rm N}$ are fairly high
in Es located in the centers of massive halos with
${\log} M_{\rm h} \sim  10^{12.5} {\rm M}_{\odot}$.
These results suggest that giant elliptical galaxies
(and cDs) in the centers of massive groups and clusters of galaxies
have higher  $S_{\rm N}$ and  $V_{\rm N}$  than those in the field
and small groups of galaxies.

\subsection{Dynamical properties}
\subsubsection{Structures}

We next investigate the half-number radii of MPCs ($r_{\rm e,mpc}$)
and MRCs ($r_{\rm e,mrc}$) for GCs within the virial radii ($r_{\rm vir}$)
of host galaxy dark matter halos: it should be stressed that
the  results below might well depend weakly
on the adopted initial distributions of GCs within their host halos
at high $z$.
The spatial distributions of MPCs have valuable information
of dark matter halos of their host galaxies (B07).
In order to avoid contributions
from intracluster and intragroup
GCs in groups and  clusters of galaxies
with $r_{\rm vir}>100$ kpc,
we estimate $r_{\rm e,mpc}$ and $r_{\rm e,mrc}$
within 100 kpc for these groups and clusters.
We define $s_{\rm eff}$ to be:
\begin{equation}
s_{\rm eff}=\frac{r_{\rm e, mrc}}{r_{\rm e, mpc}}.
\end{equation}
 We also define the power-law slope ($\gamma$) and
the coefficient ($D_{0}$)  of the following relation:
\begin{equation}
r_{\rm e, mpc}=D_{0}{M_{\rm h}}^{\gamma} .
\end{equation}

 Fig. 16 shows that $r_{\rm e, mpc}$ and $r_{\rm e, mrc}$ depend
 on $M_{\rm B}$ such that both are larger for more luminous
 galaxies.  The mean values of $r_{\rm e, mpc}$ are larger than
 10 kpc for $-22$ $<M_{\rm B}<$ $-14$. This is larger than the observed
 the half-number radius of the Galactic GC system which is about
 5 kpc (e.g., van den Bergh 2000).  Although no observational
 studies have so far investigated the correlation of $r_{\rm e,
 mpc}$ with $M_{\rm B}$, the above results imply that the
 simulation significantly overestimates $r_{\rm e, mpc}$: this
 may be true for MRCs as well.  In the present model, all GCs within
 $r_{\rm vir}$ (virial radius) of the dark matter halo of their host
 galaxy are used for estimation of $r_{\rm e, mpc}$.  This way
 of estimating $r_{\rm e, mpc}$ would contribute significantly to
 the possible overestimation of $r_{\rm e, mpc}$ of GCSs in the
 simulation.

 We find $s_{\rm eff}$ is significantly less than 1 for a wide range of
luminosities, which means that the distributions of MRCs
are more compact than those of MPCs for most GCSs.
The mean  $s_{\rm eff}$  is 0.84 in the present model,
which can not be currently compared with observations owing to the lack
of observational studies of  $s_{\rm eff}$ in galaxies.
The smaller mean value of  $s_{\rm eff}$
and the weak dependence of $s_{\rm eff}$ on $M_{\rm B}$
can be tested against future observational studies.
We also suggest that the slopes of   $r_{\rm e, mpc}-M_{\rm B}$
and   $r_{\rm e, mrc}-M_{\rm B}$ relations can be used to constrain
theoretical models of GC formation based on hierarchical clustering scenarios,
because these depend on merging histories of galactic building
blocks that form GCs.

Fig. 17 shows $r_{\rm e, mpc}$ of GCSs  correlate
with their host halo mass such that  $r_{\rm e, mpc}$
is  larger for larger  $M_{\rm h}$.
The $\chi$-square fit to the simulation data is: 
\begin{equation}
{\log}_{10} (\frac { r_{\rm e,mpc} } { {\rm kpc} }) =-0.69
+0.18{\log}_{10} (\frac{M_{\rm h}}{ {\rm M}_{\odot} })  
\end{equation}
which means $ r_{\rm e,mpc}=
0.20{( \frac{ M_{\rm h}} { {\rm M}_{\odot} }  )}^{0.18}$ kpc
(i.e., $\gamma$=0.18).
The following equation would be more useful for
making  an estimation of $M_{\rm h}$ for
a galaxy by measuring  $r_{\rm e, mrc}$ of the GCS:
\begin{equation}
 {\log}_{10} (\frac{M_{\rm h}}{ {\rm M}_{\odot} }) =
4.5+
5.2 {\log}_{10} (\frac { r_{\rm e,mpc} } { {\rm kpc} }).
\end{equation}

Since the present model 
based on a dissipationless simulation
can  overestimate $r_{\rm e, mpc}$,
it might  be better to use observations
in order to determine the zero-point
of the above power-law relation.
If we use observations of the Galactic
GCS (e.g., van den Bergh 2000 and references therein) and the total
mass of the Galaxy (Wilkinson \& Evans 1999),
then the power-law relation is 
$ r_{\rm e,mpc}=
5.0 {( \frac{ M_{\rm h}} { 2 \times 10^{12}  {\rm M}_{\odot} }  )}^{0.18}$ kpc
or 
$M_{\rm h} = 2 \times 10^{12}   
{(\frac{  r_{\rm e,mpc} } {  {\rm 5 kpc} })}^{5.2}
{\rm M}_{\odot}.  $ 
The derived $M_{\rm h}$ 
from $ r_{\rm e,mpc}-M_{\rm h}$ relations
can be compared with $M_{\rm h}$
derived from kinematics of GCSs and halo field stars 
(e.g., Romanowsky 2006).

\subsubsection{Kinematics}

 A growing number of observational data sets on the kinematical properties
of GCSs have been recently accumulated (e.g., Romanowsky 2006).
We investigate correlations
between velocity dispersions ($\sigma$) 
for MPCs (${\sigma}_{\rm mpc}$) and MRCs (${\sigma}_{\rm mrc}$)
and  $M_{\rm B}$ of their host galaxies.
We first estimate ${\sigma}_{\rm mpc}$ and   ${\sigma}_{\rm mrc}$
of  a GCS for each of the three  projections
($x$-$y$, $x$-$z$, and $y$-$z$) by using line-of-sight-velocities
of all GCs within the virial radius of the halo.
We then make an average for the three projections
and determine one-dimensional velocity dispersions of
${\sigma}_{\rm mpc}$ and   ${\sigma}_{\rm mrc}$.
We also investigate  correlations between  $M_{\rm B}$ and $s_{\rm dis}$,
where  $s_{\rm dis}$ is defined as:
\begin{equation}
s_{\rm dis}=\frac{{\sigma}_{\rm mrc}}{{\sigma}_{\rm mpc}}.
\end{equation}
 
Fig. 18 shows that both ${\sigma}_{\rm mpc}$
and  ${\sigma}_{\rm mrc}$  are higher in
more luminous  galaxies.
The essential reason for this dependence is described as follows.
Since the GCSs investigated in the present study are for galaxies
located in the centers of dark matter halos
(i.e., GCSs in satellite galaxies are not investigated),
more luminous galaxies are more likely to be embedded in
more massive halos (i.e., larger  $M_{\rm h}$).
GCs follow structures  and kinematics of underlying
dark matter halos so that the mean velocity dispersions of GCSs
are determined mainly by masses of their halos
(and by their  spatial distributions).
Therefore the velocity dispersions  of GCSs are higher
in more luminous galaxies.
The difference in $s_{\rm dis}$ for different
$M_{\rm B}$ is quite small and the mean value of $s_{\rm dis}$ is 0.98.
These results imply that velocity dispersions are not so different
between MPCs and MRCs in galaxies.

Owing to the collisionless nature of the present simulation,
the ratios of maximum rotational velocities to central velocity
dispersions ($V/\sigma$) in GCSs  appear to be 
low ($V/\sigma <0.3$) for  
most  galaxies. These results are in a striking contrast with our
previous simulations (Bekki et al. 2005)
in which some GCSs in the  remnants
of disk-disk major mergers show large $V/\sigma$ ($>0.5$).
Owing to the lack of extensive statistical  studies on
$V/\sigma$ of GCSs in galaxies,
it is not clear whether the above results are generally consistent
with observations or not.

\subsection{Mean properties}

Lastly, we briefly summarize
the key physical properties of 
GCs averaged over all GCSs in the simulation
in Table 4. The columns are the mean properties
of $z_{\rm f, mpc}$ (column 1),
$z_{\rm f, mrc}$ (2),
$f_{\rm bimo}$ (3),
$f_{\rm mrc}$ (4),
$f_{\rm y}$(5),
$S_{\rm N}$ (6),
$V_{\rm N}$ (7),
$s_{\rm eff}$ (8),
$s_{\rm dis}$ (9),
and $\epsilon$ (10).
In estimating these mean values,
we do not include galaxies  with no GCs. 
If we include galaxies with no GCs, 
the mean values would be significantly
changed, in particular,  for $S_{\rm N}$ and $V_{\rm N}$.

\begin{figure}
\psfig{file=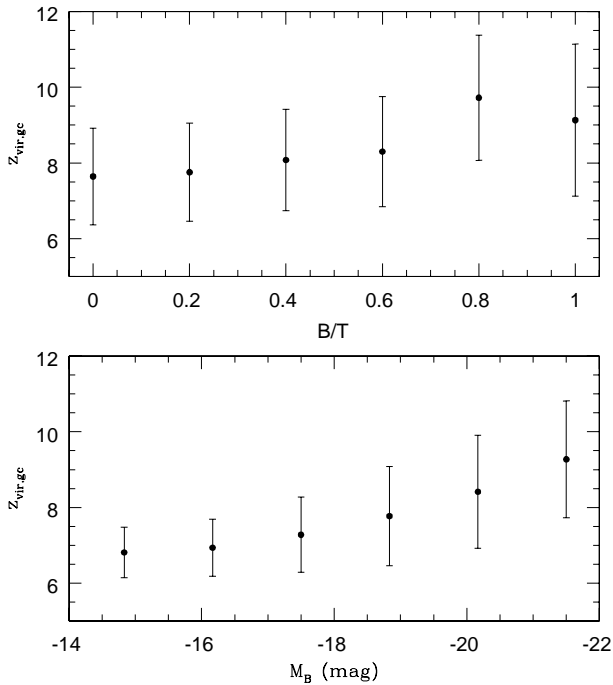,width=8.0cm}
\caption{
Distributions of the simulated galaxies with GCs in the
$z_{\rm vir,gc}-B/T$ plane (upper) and
$z_{\rm vir,gc}-M_{\rm B}$  plane (lower)
where $z_{\rm vir,gc}$ is the mean $z_{\rm vir}$  of
building blocks hosting GCs in a galaxy.
Circles show mean values in $B/T$ and $M_{\rm B}$  bins. 
The $1\sigma$ dispersion
for each  bin is shown by an error bar.
No galaxies have $z_{\rm vir,gc} < 6$, because
formation of GCs in the present model is truncated for galaxies
with  $z_{\rm vir}<6$.
}
\label{Figure. 19}
\end{figure}

\begin{figure}
\psfig{file=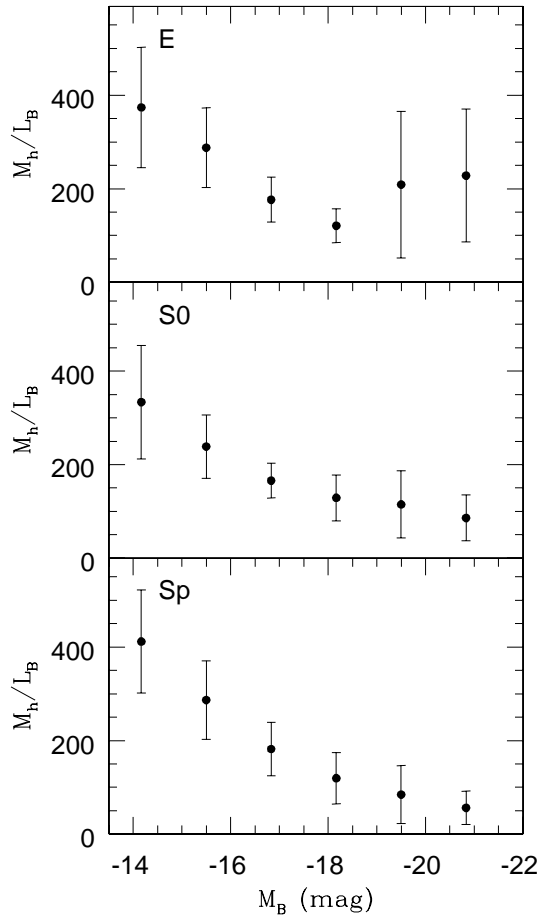,width=7.0cm}
\caption{
Distributions of the simulated galaxies with GCs in the
$M_{\rm h}/L_{\rm B}-M_{\rm B}$  plane,
where $L_{\rm B}$ is the total $B-$band luminosity of a galaxy, 
for E (top),
S0 (middle), and
Sp (bottom).
Circles show mean values in $M_{\rm B}$  bins. 
The $1\sigma$ dispersion
for each  bin is shown by an error bar.
}
\label{Figure. 20}
\end{figure}

\section{Discussion}

\subsection{Bimodal colour distributions in GCSs}

The origin of the observed bimodal colour distributions of GCSs in
elliptical galaxies have long been discussed in the context
of different formation scenarios of elliptical galaxies,
such as two-phase collapse at high redshift (Forbes et al. 1997),
accretion and stripping of low-mass galaxies with GCs (C\^ote et al. 1998),
and gas-rich major mergers (Ashman \& Zepf 1992).
B02 was the first to investigate the bimodal colour distributions
in a quantitative manner based on the results of a semi-analytic
model of galaxy formation.  B02 showed that the observed
bimodality can be reproduced, if  MRCs are formed during
dissipative merger events at high redshift and that the formation
of MPCs are truncated at $z \sim 5$.  B02 however did not
constrain the truncation mechanism.

Since age differences between MPCs and MRCs are rather small ($<1$ Gyr)
in the present simulation,
we can discuss the simulated MDFs of GCSs in terms of the colour bimodality
of GCSs.
In the present model,  all GCs (i.e., both MPCs and MRCs) in a  galaxy
originate from low-mass building blocks at $z$ $>3$,
whether they are isolated or in merging galaxies.
Therefore the origin of MRCs in a galaxy is not necessarily associated with
the past major merger events of a galaxy. 
The peak metallicity  in the MDF for MPCs (MRCs) in  a galaxy
reflects the highest GCFR in the galaxy's building blocks
that have stellar metallicities lower (higher) than $-1$ at high $z$.
The mean metallicity of MPCs (MRCs)  in a galaxy is determined by
the mean stellar metallicity of the more metal-poor  building blocks
with ${\rm [Fe/H]} <-1$ (${\rm [Fe/H]} \ge -1$).
Although MRCs are formed later in more metal-rich building blocks of galaxies,
there is only a slight difference in formation redshifts between MPCs and MRCs.
The major merger events in which MRCs are formed are typically $z$ $>3$
whereas the last major merger events
which determine the final morphological types of galaxies 
can happen later at lower redshifts.
Therefore the last major merger events responsible for elliptical galaxy formation
are not  necessarily associated with the formation of  MRCs 
,  which is in contrast to 
the scenario proposed by Ashman \& Zepf (1992) but consistent
with the so-called `damp' merger interpretation of Forbes et
al. (2007).

In order to show the key requirements for the formation of bimodal
MDFs in GCSs, 
we have run two comparative models with $\alpha=0.1$ and $\beta=0.05$
and  $\alpha=0.02$ and $\beta=0.25$ (see section 2.3).
For these models,  the dependences of the GCFR on $f_{\rm m}$ and $f_{\rm
g}$ are assumed to be very weak.
We find that GCSs in these models show much reduced MDF
bimodality in comparison  with the present standard model
with  $\alpha=0.02$ and $\beta=0.05$ shown in Table 1
(the results of the models are shown in the Appendix B).
These results clearly show that a strong enhancement in the GCFR
during violent merging (in particular, major merging)
and gas-rich phases (i.e., at high redshifts) is important
for the formation of the bimodality.

The origin of the metal-poor GC peaks is due mainly to the strong
enhancement of the  GCFR in gas-rich and high gas-density
building blocks at high-$z$ whereas that of the metal-rich
ones is due mainly to the strong enhancement of the GCFR in more metal-rich
and gas-rich building blocks that experience violent merging
at somewhat later epochs.
The truncation of GC formation via reionization is necessary
for the present model in order 
to prevent too many GCs from forming in low-mass galaxies.
The epoch of reionization, however, appears not to determine the 
presence of bimodality itself.  Although differences
in the merging histories of halos 
between different galaxies can introduce a
diversity in GCS properties,  such differences appear to have
no {\it  direct} link with  the resulting bimodal  MDFs of the GCSs.

As shown in Fig. 8, not all of galaxies have GCSs with bimodal MDFs.
In order to understand why these galaxies have GCSs without 
bimodality, we have investigated formation histories of GCs
in these galaxies (the results for some galaxies are shown
in the Appendix B).  We find that these galaxies have only either metal-poor
or metal-rich peaks in their MDFs owing to single burst
events of GC formation. 
Galaxies with no, or little, enhancement in the GCFR for metal-poor GCs
and strong enhancement of the GCFR for metal-rich GCs
have no clear bimodality in their MDFs. 
On the other hand, those with strong enhancement of the GCFR
for metal-poor GCs at high $z$ and 
no enhancement of the GCFR for metal-rich GCs
do not show the bimodality either.
The efficient formation of MPCs is highly unlikely for
galaxies with lower $z_{\rm vir}$ whereas 
MRCs are highly unlikely to be
formed in gas-poor merging at later epochs.
The galaxies having GCSs without bimodal MDFs thus have
either lower $z_{\rm vir}$ or few  events of gas-rich 
merging and accretion of more metal-rich galactic building blocks.

The present model has shown that
a significant fraction ($f_{\rm bimo} \sim 0.2$) of very late-type spirals with
small or no  bulges ($B/T<0.05$) have bimodal MDFs in their GCSs.
This is because more metal-rich (${\rm [Fe/H]} \ge -1$)
building blocks of late-type spiral galaxies  can form GCs at high $z$
which are later accreted during mergers. 
Therefore the origin of MRCs (and thus bimodal MDFs)
in late-type spiral galaxies are not necessarily associated with
the formation of galactic bulges via early major merger events
in the present model (see B02 for a brief discussion on this issue).
If most 
bulge-less spirals are found to contain very few MRCs (as appears to be the case for 
M33; Chandar et al. 2006) 
then the model proposed here may require significant alteration. 

\begin{table*}
\centering
\begin{minipage}{185mm}
\caption{  Qualitative assessment of model successes and failures.}
\begin{tabular}{cccc}
Items
& {Consistency  \footnote{ $\bigcirc$ ($\times$) means
that simulations are broadly consistent (inconsistent) with observations.
 $\bigtriangleup$ means  that simulations are only partly
consistent with observations.}}
& Comments \\
Ages 
&  $\bigcirc$
&  Data indicate very old ages for both MRCs and MPCs. \\
Bimodal MDFs 
&  $\bigcirc$
&  Requires predicted MDFs in observed colours. \\
$f_{\rm bimo}-M_{\rm B}$  
& $\bigcirc$
&  Broadly consistent but data very limited.  \\
$f_{\rm mrc}-M_{\rm B}$
&  $\bigcirc$
& Consistent with data for luminous galaxies. \\
$f_{\rm y}$
&  $\bigcirc$
&  Broadly consistent but data very limited. \\
$M_{\rm B}-{\rm [Fe/H]}$
&  $\bigcirc$
& Less consistent for MPCs. \\
$S_{\rm N}-B/T$
&  $\bigcirc$
& Consistent in terms of higher $S_{\rm N}$ in Es. \\ 
$S_{\rm N}-M_{\rm B}$
& $\bigtriangleup$
&  U-shape of data not clearly seen.\\ 
$V_{\rm N}-M_{\rm h}$
&   $\bigtriangleup$
&  Data show a flatter slope. \\
$r_{\rm e, mpc}$
& $\bigtriangleup$
&  Data for the Galaxy is much smaller.\\ 
\end{tabular}
\end{minipage}
\end{table*}

\subsection{Specific Frequencies $S_{\rm N}$ and $V_{\rm N}$}

Previous observations have revealed that the $S_{\rm N}$ of GCSs
depends on the Hubble type and luminosity of the host galaxy
(e.g., Harris 1991; but see also Spitler et al. 2007 for an
alternative view).  Although the observed trends
of $S_{\rm N}$ with galactic properties (e.g., the bimodality in
the $M_{\rm V}-S_{\rm N}$ relation) are well reproduced by B06,
other key observations of $S_{\rm N}$ have not been discussed.
In the present study, we have shown that (i) $S_{\rm N}$ of
GCSs in luminous galaxies with $M_{\rm B}<-19$ mag are
significantly higher in early-type (E/S0) galaxies than late-type
(Sp) ones, (ii) S0s typically have higher $S_{\rm N}$ than
spirals, and (iii) low-luminous galaxies with $M_{\rm B}> -15$
mag have higher $S_{\rm N}$, regardless of their Hubble types.
These results are qualitatively consistent with previous
observations by Harris (1991), Forbes (2005), and
Arag\'on-Salamanca et al. (2006).  So far we have not
discussed the physical reasons for the above dependences of
$S_{\rm N}$ on galactic properties.

Fig.  19 shows that the mean values  ($z_{\rm vir,gc}$)  
of virialization  redshifts ($z_{\rm vir}$)
of galactic
building blocks that form GCs 
are significantly higher in early-type galaxies with $B/T>0.8$
than in late-type ones.
This means that GCFRs in building blocks of early-type
galaxies are higher owing to their higher mass-densities
with higher $z_{\rm vir}$. 
Because GCFRs depends strongly on $z_{\rm vir}$
through the term of $F_{\rm z}$ (see equation 6) the higher $z_{\rm vir,gc}$
in early-type galaxies is one reason for the observed higher $S_{\rm N}$
of their GCSs.
Furthermore Fig. 19 shows that $z_{\rm vir,gc}$ is higher in more luminous
galaxies, which can explain why more luminous early-type galaxies
with $M_{\rm B}<-19$  mag are more likely to have higher
$S_{\rm N}$ in comparison with less luminous ones with
$-19.0$  $ < M_{\rm B} < -17$ mag  seen in Fig. 12.

Fig. 19 shows that low-luminosity galaxies do not have higher
$z_{\rm vir,gc}$, which implies that higher $S_{\rm N}$ in
low-luminosity galaxies seen in Fig. 9 can not be understood in
terms of $z_{\rm vir,gc}$.  B06 suggested that the origin of the
observed higher $S_{\rm N}$ in low-luminosity galaxies is due to
higher mass-to-light-ratios ($M/L \approx M_{\rm h}/L_{\rm B}$)
of these galaxies.  Fig. 20 shows that $M_{\rm h}/L_{\rm B}$
steeply depends on luminosity in the sense that $M_{\rm
h}/L_{\rm B}$ is higher in less luminous galaxies.  Given that GC
numbers per unit halo mass ($V_{\rm N}$) in galaxies only weakly
depend on the halo masses of the galaxy (B06; see also Fig.13
in the present paper), the above result implies that the higher
$S_{\rm N}$ in low-luminosity galaxies are due mainly to their higher
$M_{\rm h}/L_{\rm B}$ ratios owing to the relation  $S_{\rm N}=V_{\rm
N} (\frac{ M_{\rm h} }{ L_{\rm B} })$.

\subsection{Metallicity-luminosity  relations}

Although B02 predicted that the mean colours of both MPCs and MRCs
only weakly correlate with the total luminosities of their host galaxies,
these predictions are inconsistent 
with latest observations by Strader et al. (2004) and P06.
The present model has successfully reproduced reasonably well,
the observed ${\rm [Fe/H]} - M_{\rm B}$ relations
for MRCs and all GCs.
The derived  slope in
the ${\rm [Fe/H]} - M_{\rm B}$ relation for MPCs in the present model
is, however,  not consistent with the observed one.
The differences between
B02 and the present study are caused by  the differences
in the adopted models of GC formation
between  these two studies.
Thus more sophisticated models of GC formation in galaxies
at high $z$ need to be developed so that all 
three  ${\rm [Fe/H]} - M_{\rm B}$ relations  (MPCs, MRCs, and all GCs)
can successfully reproduce the observations. 

B07 showed that if $z_{\rm trun} \approx 10$, the observed ${\rm
[Fe/H]} - M_{\rm B}$ relation for MPCs is better reproduced.
However the simulated $\epsilon$ (i.e., the formation efficiency
of GCs) in the model with $z_{\rm trun} =10$ in the present study
is too small ($0.00077$) to be consistent with the observed one
by McLaughlin (1999). The present model seems to have
difficulties in explaining self-consistently both the observed
${\rm [Fe/H]} - M_{\rm B}$ relation for MPCs and $\epsilon$.  One
possible way to solve this problem would be to investigate models
with higher $z_{\rm trun}$ (=$6-10$) in which $C_{\rm eff}$
depends more strongly on $z$ than in the present model so enough GCs
can be formed at $z=6-10$.  It is, however, unclear
whether the models with higher $z_{\rm trun}$ and $C_{\rm eff}$
can reproduce other key observations such as the $S_{\rm N}-M_{\rm
B}$ relation.

\subsection{Success and failures of the present model}

Although we have so far presented variously different physical properties
of GCSs in galaxies,
only some of them can be directly compared with observations: for example,
the simulated
$f_{\rm mrc}-M_{\rm B}$ relation can be compared with observations by P06
whereas the simulated  $s_{\rm dis}-M_{\rm B}$ one can not owing to
the lack of observational data.
It is, however, important to check whether the simulated properties
are consistent with the corresponding observations.
In Table 5 we summarise what we regard as the relative successes
and failures of the current model in a qualitative sense. For
any given physical property  there are
subtle, and sometimes large, differences between the
observational data and the corresponding property
predicted by the model.
The references of these observational properties are already given 
in the Introduction section \S 1 (e.g., Brodie \& Strader 2006)
and some of the observational data are presented within  figures
of this paper.
Less consistent results (e.g., the $M_{\rm B}-{\rm [Fe/H]}$
relation for MPCs) imply  that a more sophisticated model for
GC formation would be required for more successful modeling.

\section{Conclusions}

We have investigated the structural, kinematical, and chemical properties
of globular cluster systems (GCSs) in galaxies
in a self-consistent manner based on
high-resolution cosmological N-body simulations combined with
a semi-analytic model  of galaxy 
and globular cluster (GC) formation.
We have adopted a number of assumptions on formation efficiencies
of GCs which depend on the physical properties of their host galaxies
(e.g., gas mass fraction).
We investigated
correlations between
physical properties of GCSs and those of their host galaxies
for $\sim 10^5$
simulated central halo galaxies for MPCs with ${\rm [Fe/H]} <-1$
and MRCs with ${\rm [Fe/H]} \ge -1$.

We find:
(1) The majority ($\sim$ 90\%) of GCs currently
in halos of galaxies  are formed in low-mass galaxies at  $z > 3$
with mean formation redshifts of MPCs and MRCs being 5.7 and 4.3,
respectively. This corresponds to 12.7 and 12.3 Gyrs in lookback
time. 
The majority of MPCs are formed in low-mass galaxies  that are virialized
well before reionization ($z>6$) and thus have higher mass densities.
MRCs are formed slightly later not only within  high $z$  major mergers between
high-density galaxies but also within 
gas-rich isolated ones.

(2) About 50 \% of galaxies with GCs show clear bimodalities in
their MDFs, 
though less luminous galaxies with $M_{\rm B}$ fainter than $-17$ 
are much less likely  to show the bimodalities owing to
no or few MRCs. The 
age differences between MPCs and MRCs are quite small ($<$ 1 Gyr)
in most galaxies.
The origin of the simulated bimodality in MDFs is  due to
strong dependences of GCFE on $f_{\rm g}$
and $f_{\rm m}$ (i.e., higher GCFE in more gas-rich high-z
galaxies and in mergers with larger mass ratios of the merging two
galaxies).

(3) The number fractions of MRCs ($f_{\rm mrc}$) 
range from 0 to
almost 1 with an average of $0.4$.
The $f_{\rm mrc}$ in galaxies does not depend on their  Hubble type
(i.e., bulge-to-disk-ratios), which implies that
the formation of MRCs is not  necessarily associated with
bulge formation.
The $f_{\rm mrc}$  are likely to be smaller for less luminous galaxies.

(4) The $S_{\rm N}$ of GCSs  are typically
higher in ellipticals than in spirals,
and in low-luminosity galaxies with $M_{\rm B}>-15$ 
regardless of their Hubble types.
The mean  $S_{\rm N}$ for Sp, S0, and E populations
are 1.8, 2.0, and 4.0, respectively.

(5) The number of GCs per halo mass ($V_{\rm N}$) does  
not depend as strongly on
the luminosity or the Hubble type of the host galaxy
as $S_{\rm N}$ does,
which suggests that the GC number per unit mass
is similar between different galaxies.
$V_{\rm N}$ is, however, likely to be higher
in luminous ellipticals with $M_{\rm B}<-20$.

(6) Although there are no significant
differences in $V_{\rm N}$
between spirals and S0s for luminous galaxies
($M_{\rm B} <-20$),
S0s are more likely to have higher 
$V_{\rm N}$ 
than spirals for less luminous galaxies with $M_{\rm B} > -18$.
These results suggest that 
only  luminous S0s with moderately
high $S_{\rm N}$ can be transformed spirals via truncation
of star formation and the resultant disk fading,
because $V_{\rm N}$ does not change during disk fading.

(7) The mean metallicities of GCs ([Fe/H]) for MPCs and MRCs
depend on $M_{\rm B}$ of their host galaxies
such that they are higher in more luminous galaxies,
though the dependence for MPCs is weak.
Although the observed correlation of $M_{\rm B}-{\rm [Fe/H]}$
for MRCs 
can be well reproduced by the present model,
that for MPCs 
is not consistent with the simulated one.
The observed correlation between
the mean metallicities of all GCs and  $M_{\rm B}$ 
can be well reproduced by the present model.

(8) Spatial distributions of MRCs are more compact than those of MPCs
with $r_{\rm e,mrc} \sim 0.84 r_{\rm e, mpc}$ .
The $r_{\rm e,mpc}$ depends strongly on halo mass $M_{\rm h}$ such that
$r_{\rm e,mpc} \propto {M_{\rm h}}^{0.18}$
(or $M_{\rm h} \propto {r_{\rm e}}^{5.2}$)
which implies that $r_{\rm e,mpc}$ can be used for estimating
the total masses of dark matter halos.

(9) There is no significant difference in velocity dispersion
between MPCs and MRCs. Velocity dispersions are larger
in more luminous galaxies both for MPCs and MRCs regardless of
their Hubble types.

(10) The physical properties of GCSs such as MDFs, $f_{\rm mrc}$,
$S_{\rm N}$, and $V_{\rm N}$ are quite diverse between different galaxies,
depending on the virialization
redshifts of their building blocks and subsequent merging, star formation,
and chemical evolution histories of the building blocks.

\section*{Acknowledgments}
We are  grateful to the anonymous referee for valuable comments,
which contribute to improve the present paper.
K.B. and  D.A.F. acknowledge the financial support of the Australian Research 
Council throughout the course of this work.
H.Y. acknowledges the support of the research fellowships of the Japan
Society for the Promotion of Science for Young Scientists (17-10511).
MN was supported by the Grant-in-Aid for the Scientific Research Fund
(18749007) of the Ministry of Education, Culture, Sports, Science and
Technology of Japan and by a Nagasaki University president's Fund
grant.
We are  grateful to Lee Spitler for providing observational
data and discussing the present results with us.
The numerical simulations reported here were carried out on 
Fujitsu-made vector parallel processors VPP5000
kindly made available by the  Center for Computational Astrophysics (CfCA)
at National Astronomical Observatory of Japan (NAOJ)
for our  research project why36b and uhy09a.

\appendix

\section{Individual cases}

\begin{figure}
\psfig{file=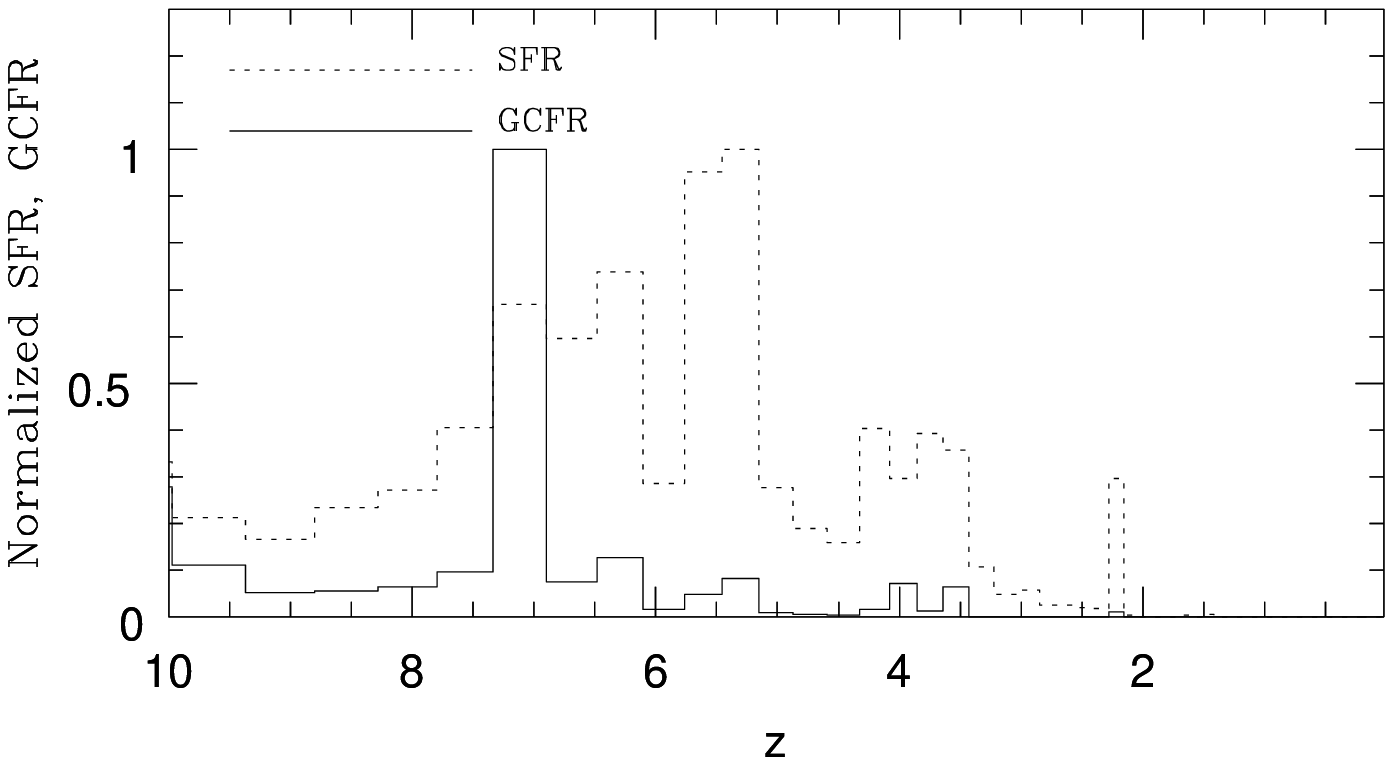,width=8.0cm}
\caption{
SFR (dotted) and GCFR (solid) as a function of $z$ in an early-type galaxy
(G1)  located
in the central region of a massive cluster of galaxies
with $M_{\rm h}=6.3 \times 10^{14} {\rm M}_{\odot}$.
G1  has a $B/T=1.0$, $M_{\rm B}= -20.7$ mag,
and   $M_{\rm V}= -21.5$ mag.
For comparison, the SFR and GCFR normalized by their maximum values
are shown.
}
\label{Figure. 21}
\end{figure}

\begin{figure}
\psfig{file=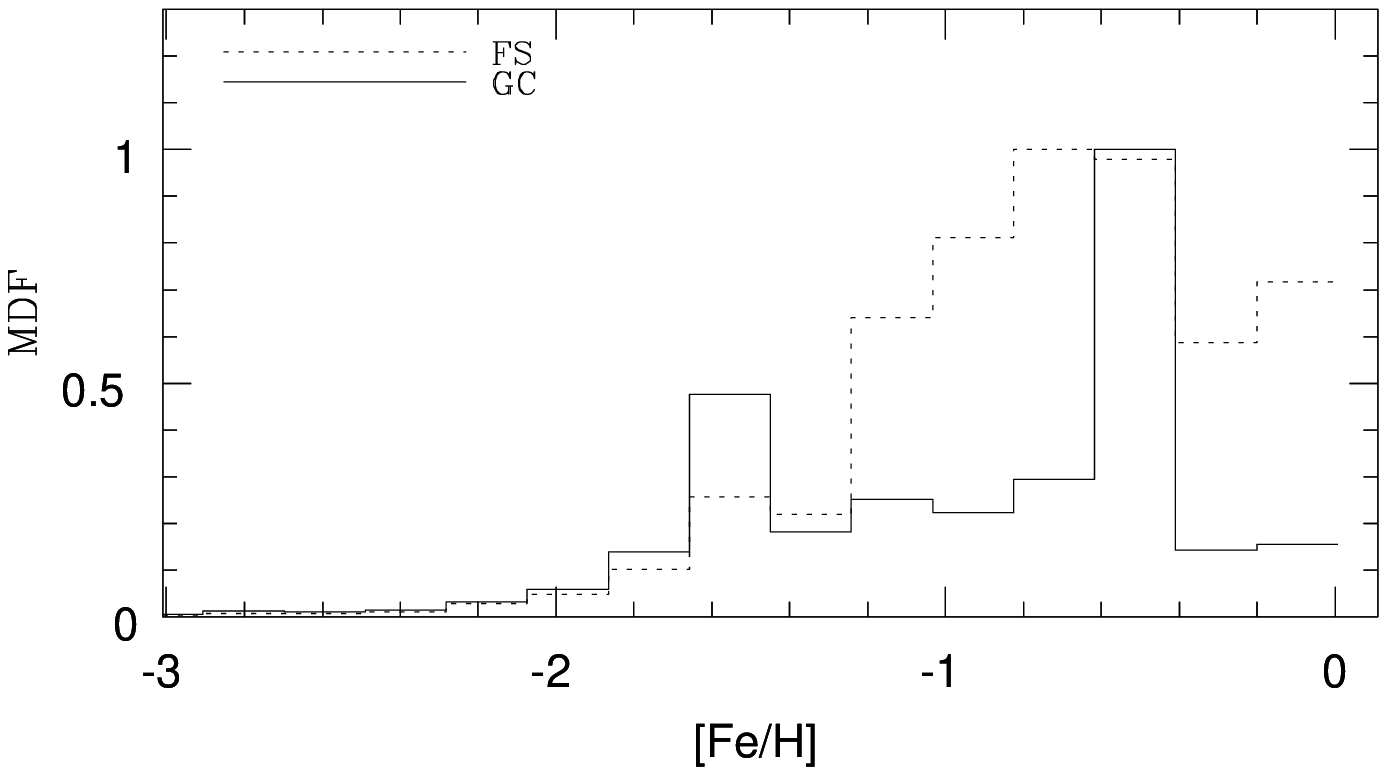,width=8.0cm}
\caption{
Solid and dotted lines represent
the MDF of GCs in the galaxy G1
and that of all field stars (FS) 
that were formed in  galaxies previously hosting the GCs
of G1.
Note that only the GCS shows a clear bimodal MDF.
}
\label{Figure. 22}
\end{figure}

\begin{figure}
\psfig{file=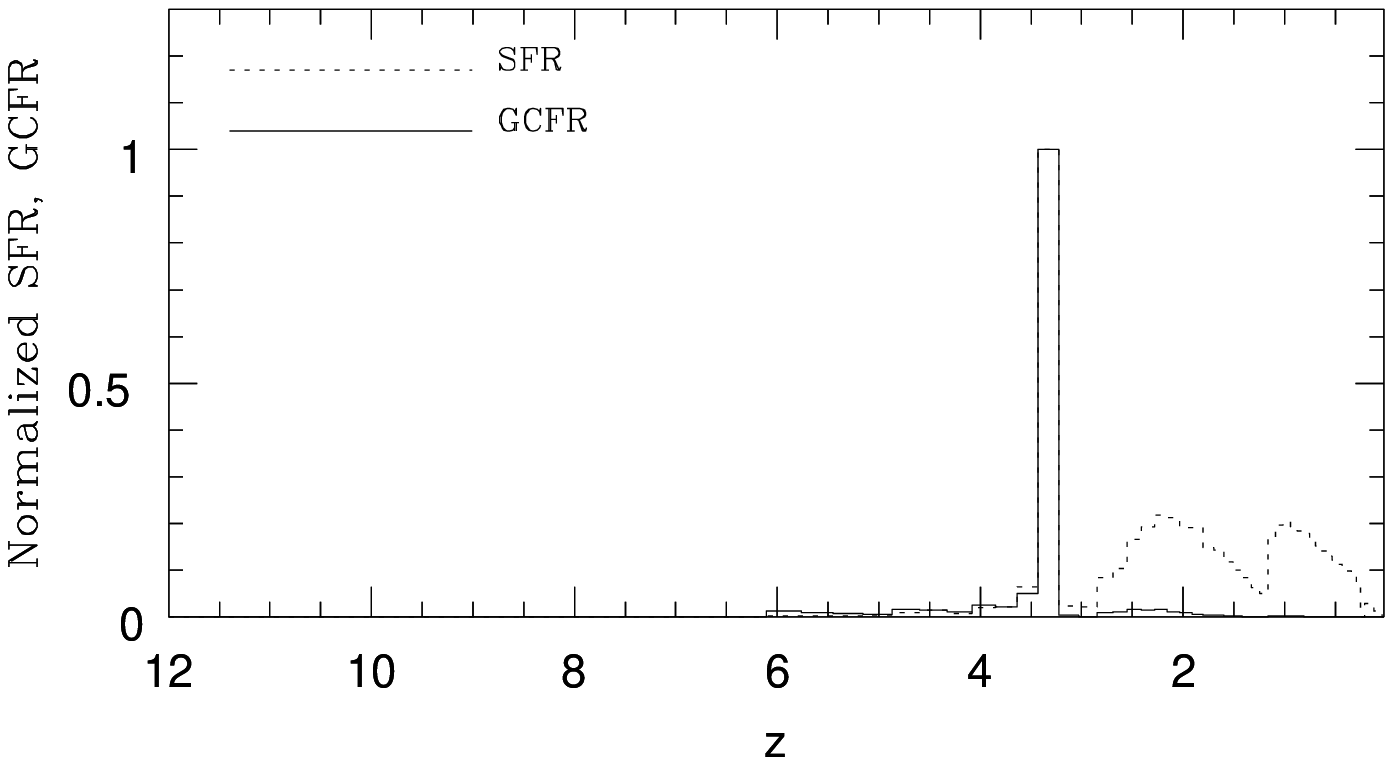,width=8.0cm}
\caption{
The same as Fig. A1  but for a spiral galaxy G2 
with $M_{\rm h}=4.1 \times 10^{11} {\rm M}_{\odot}$.
G2  has a $B/T=0.13$, $M_{\rm B}= -18.9$ mag,
and   $M_{\rm V}= -19.6$ mag.
}
\label{Figure. 23}
\end{figure}

\begin{figure}
\psfig{file=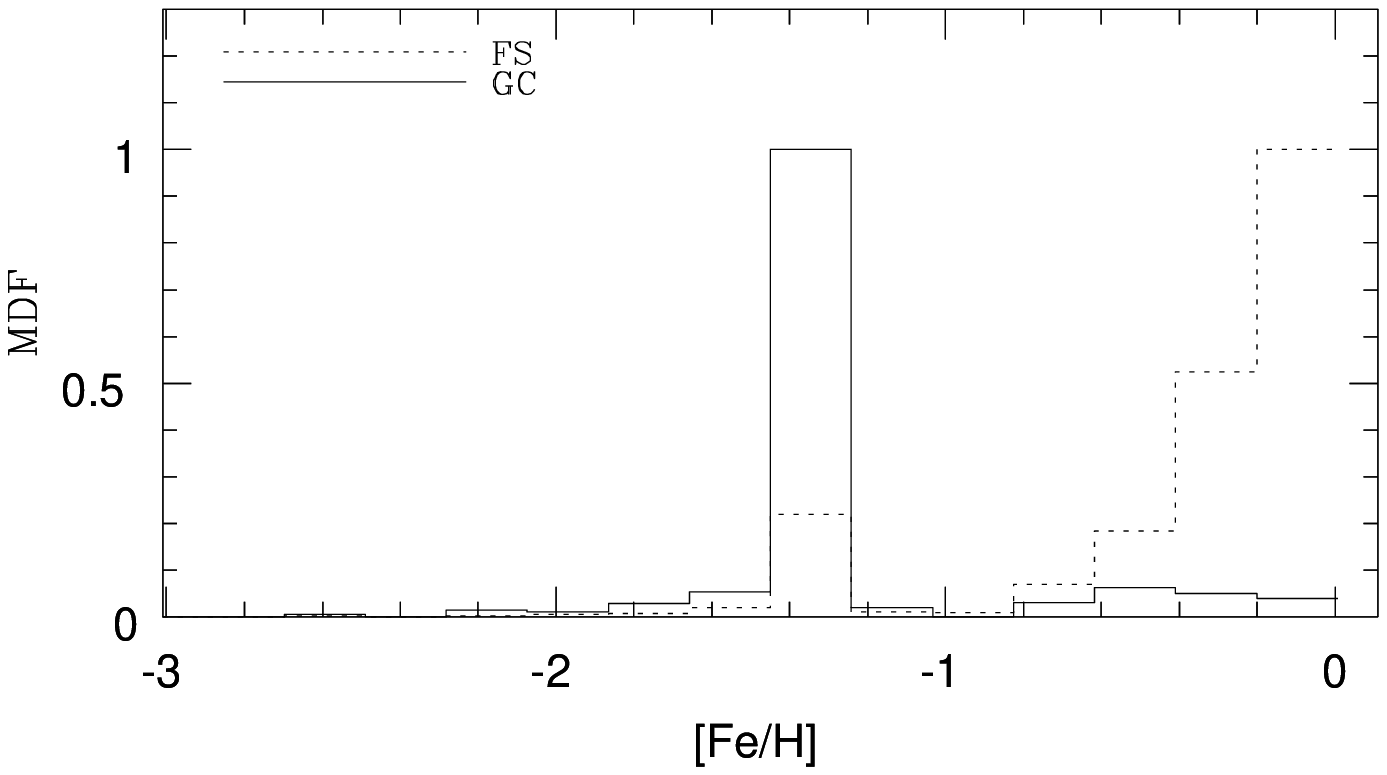,width=8.0cm}
\caption{
The same as A2  but for a spiral galaxy G2.
}
\label{Figure. 24}
\end{figure}

\begin{figure}
\psfig{file=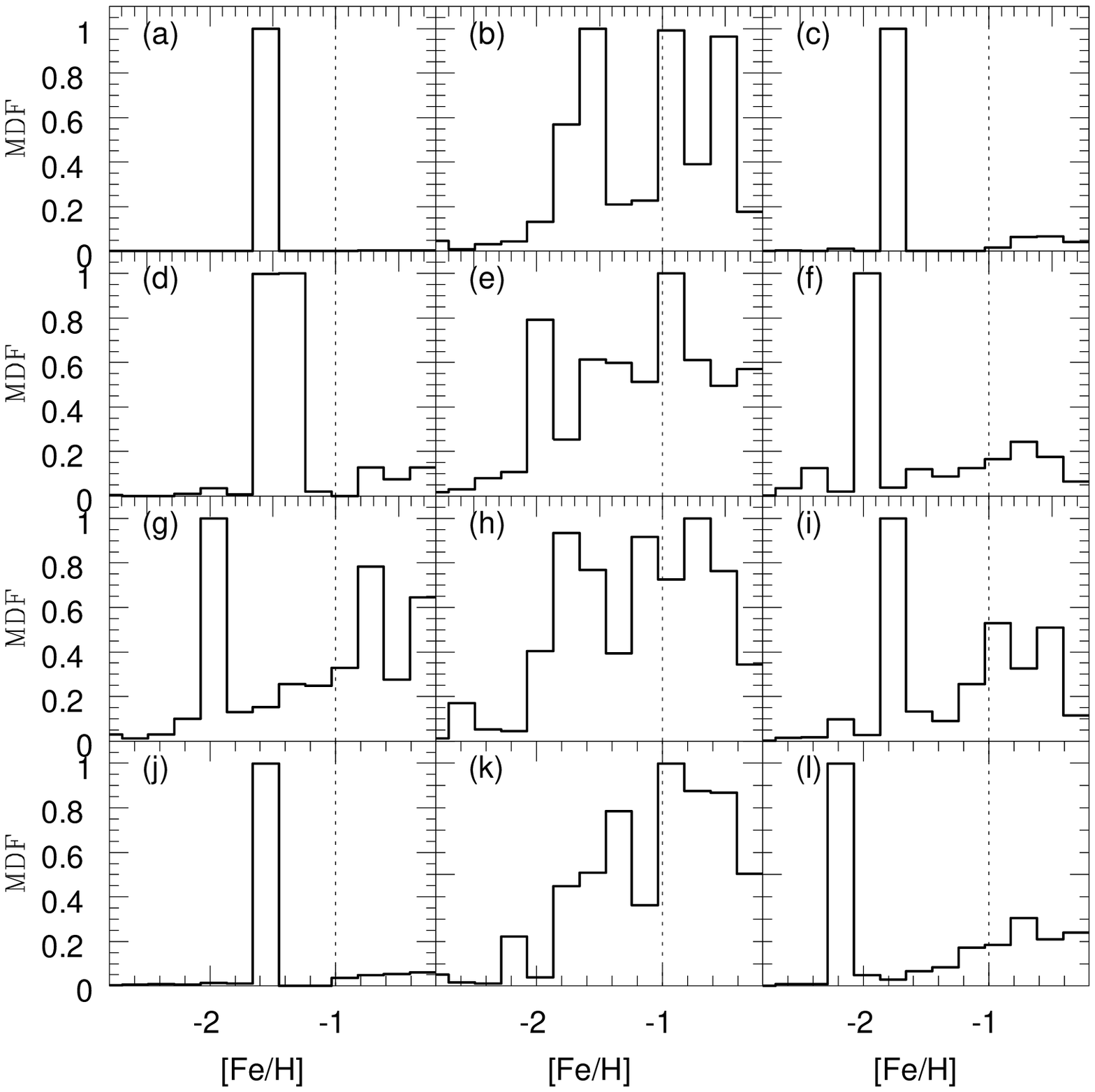,width=8.0cm}
\caption{
A collection of MDFs of 16 GCSs in spiral galaxies with $B/D < 0.68$
and with different $M_{\rm h}$. The 
16 spiral galaxies are selected
from those  embedded in dark matter halos with
$11.5 \le {\log}_{10} (\frac{ M_{\rm h} } { {\rm M}_{\odot} }) \le 12.0$.
Here  GCSs are selected such that just one GCS can be allocated in
each of 16 equally spaced mass bins in $ {\log}_{10}  M_{\rm h}$
over the above mass range.
GCSs in lower  $M_{\rm h}$
are given earlier  characters in alphabetical orders:
the GCSs in the lowest  and highest $M_{\rm h}$ are labeled as (a) and (l),
respectively.
The threshold  metallicities between  MPCs and  MRCs
(i.e., ${\rm [Fe/H} =-1$)  are shown
by dotted lines for comparison. 
}
\label{Figure. 25}
\end{figure}

\begin{figure}
\psfig{file=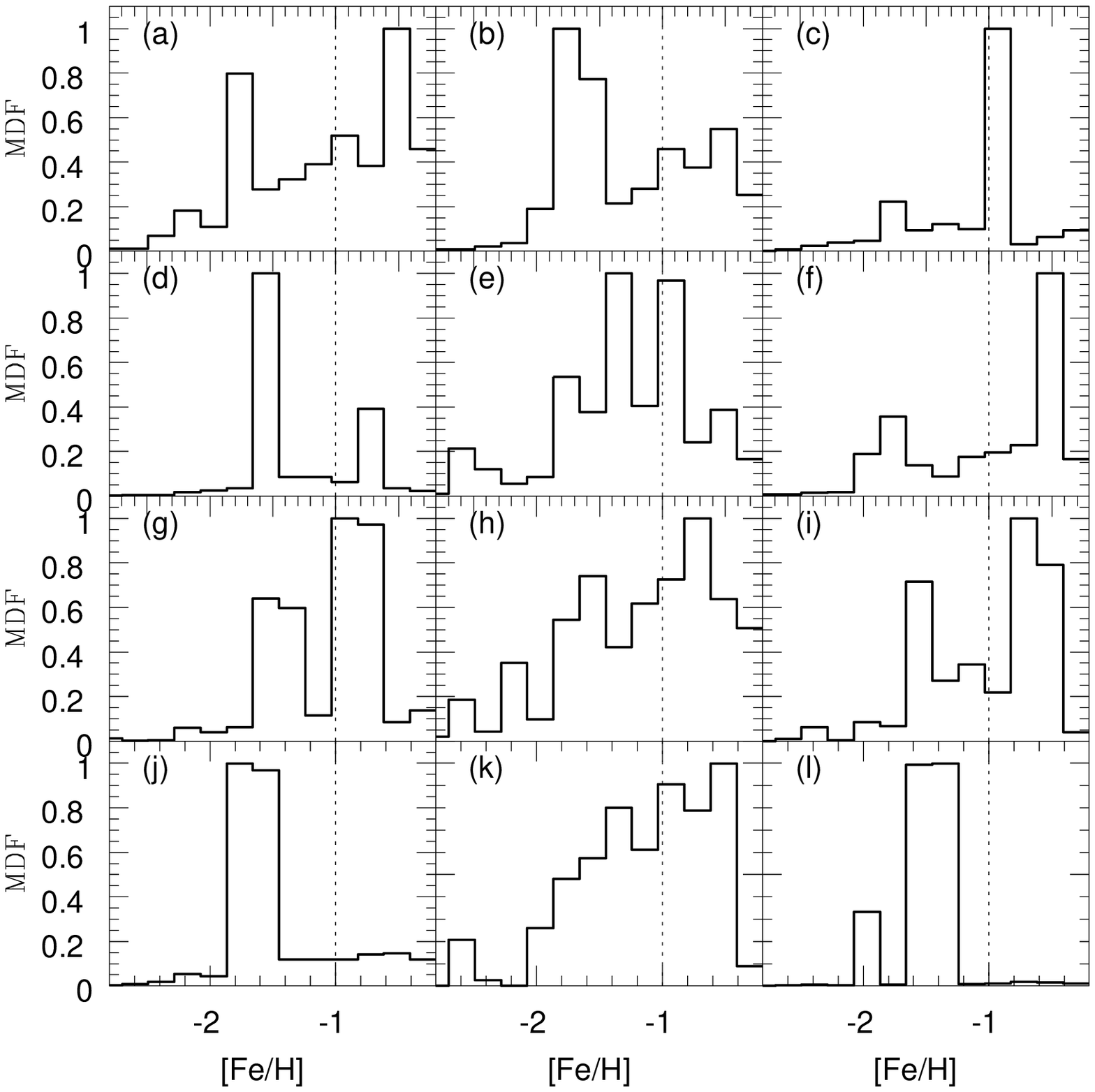,width=8.0cm}
\caption{
The same as Fig. A5 but for
16 GCSs in early-type (E and S0)
galaxies with $B/D \ge 0.68$
and with different $M_{\rm h}$.
The 16 early-type  galaxies are selected
from those embedded in dark matter halos with
$11.5 \le {\log}_{10} (\frac{ M_{\rm h} } { {\rm M}_{\odot} }) \le 13.0$.
}
\label{Figure. 26}
\end{figure}

Here we briefly discuss some results of GCSs in individual model galaxies
(G1 and G2).
Fig. A1 shows an elliptical  galaxy 
(G1) at the center of a massive
cluster of galaxies with $M_{\rm h}=6.4 \times 10^{14} {\rm M}_{\odot}$
which has two strong peaks in the GCFR at high redshifts, $z\sim 7$ and $\sim 10$.
On the other hand,  SFRs of building blocks of G1 
(including G1) show multiple
peaks at $2<z<6$, which means that formation of field stars can
continue to be quite active till relatively recently.
For G1,
both the vast majority of field stars and GCs are formed till $z=2$.
Fig. A2 shows that although the MDF of GCs shows clearly a bimodality,
that of field stars does not.
Given that some fraction of {\it these field stars
located in the outer part of G1} can finally become halo field
stars in G1 in the present model,
the above result suggests that stellar halos of ellipticals
can have unimodal MDFs as opposed to bimodal ones seen in their GCSs.

Fig.  A3 shows that in a  simulated late-type spiral (G2), 
the vast majority
of GCs are formed before $z=3$ whereas field stars can continue to
form till $z=0$, which is in significant contrast
with the results for the cluster ellipticals shown in Fig. A1.
It should be stressed here that although building blocks of G2
are virialized before $z=6$, 
intensive GC formation happens around $\sim 3$.
This means that GC formation in G2 is triggered
by merging of the building blocks (with $z_{\rm vir}>6$)
at relatively later redshifts.
As shown in Fig. A4,
G2 also shows
a lower peak metallicity for MRCs
(${\rm [Fe/H]} \sim -0.6$)
and a smaller number of MRCs.
The peak metallicity for metal-poor field stars is almost coincident
with that for MPCs, which suggests that
the MDFs are not so different between MPCs and halo field stars
in G2. It should be stressed here that
some galaxies with $M_{\rm B}$,  $M_{\rm h}$,
and Hubble types similar to G2 have shapes of MDFs in GCSs
significantly different from that of G2. The variety 
in merging histories between different galaxies
can cause these differences in MDFs.

 Fig. A5 shows that the MDFs in spirals embedded in dark matter halos
with different halos masses have variously different MDFs:
spirals labeled as (a), (c),  and (j)  show remarkably strong peaks
for MPCs
owing to a larger lumber of MPCs
whereas those labeled as (b), (e),  (h), and (i) show clearly
multiple (triple) peaks in MDFs.
Spirals labeled as (f) and (l) can be considered to be showing
clearly bimodal MDFs in their GCSs.
The relative contributions of MRCs to MDFs are clearly different
between different GCSs in these 16 spirals.
Fig. A6 shows that MDFs in ellipticals embedded in dark matter halos
with different halos masses have variously different MDFs:
a larger number of GCSs appear to have bimodal MDFs in ellipticals
than in spirals.
These diversities in the shapes of MDFs in galaxies
with different luminosities and Hubble types  reflect the diversities
in merging  and star formation histories between the galaxies.

\section{Comparative models}

\begin{figure}
\psfig{file=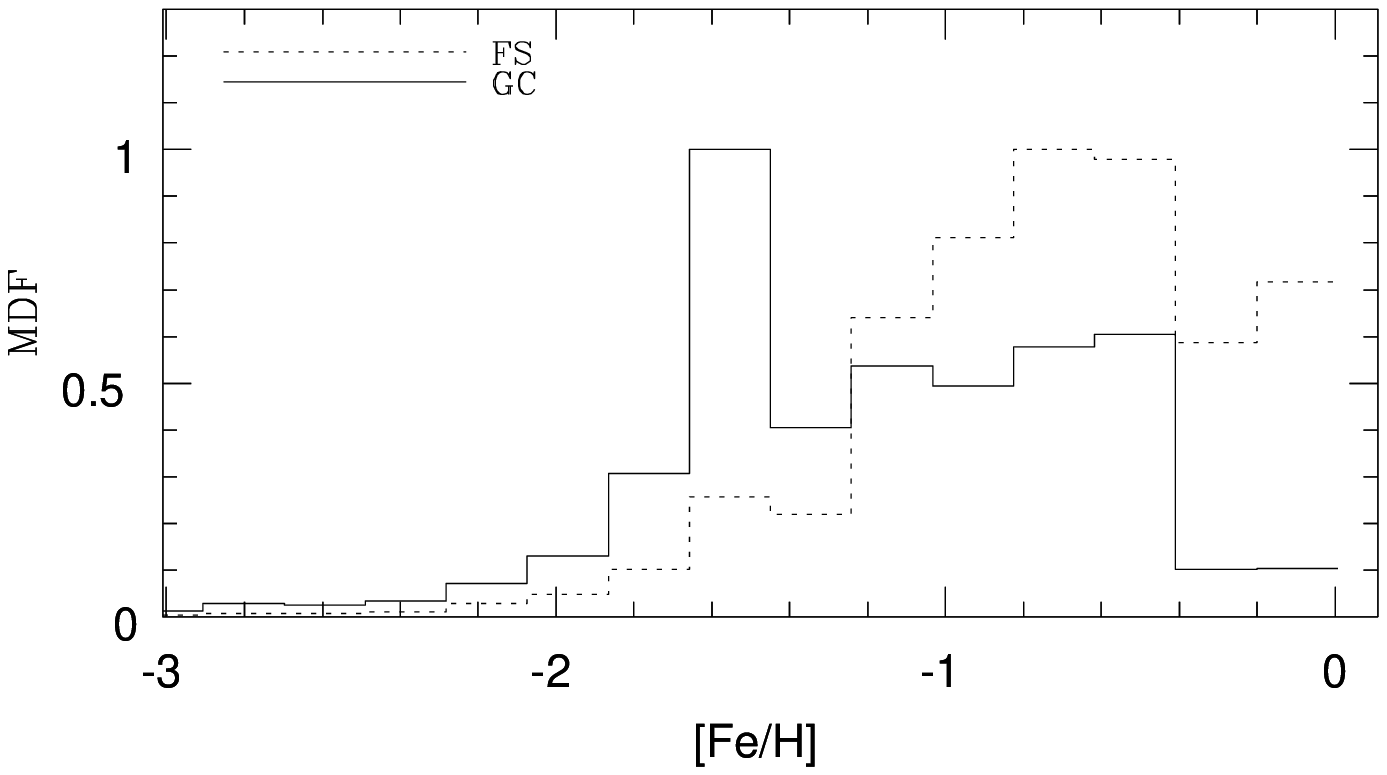,width=8.0cm}
\caption{
The same as Fig. A2  but for the model with $\alpha = 0.1$ 
and $\beta =0.05$.
}
\label{Figure. 27}
\end{figure}

\begin{figure}
\psfig{file=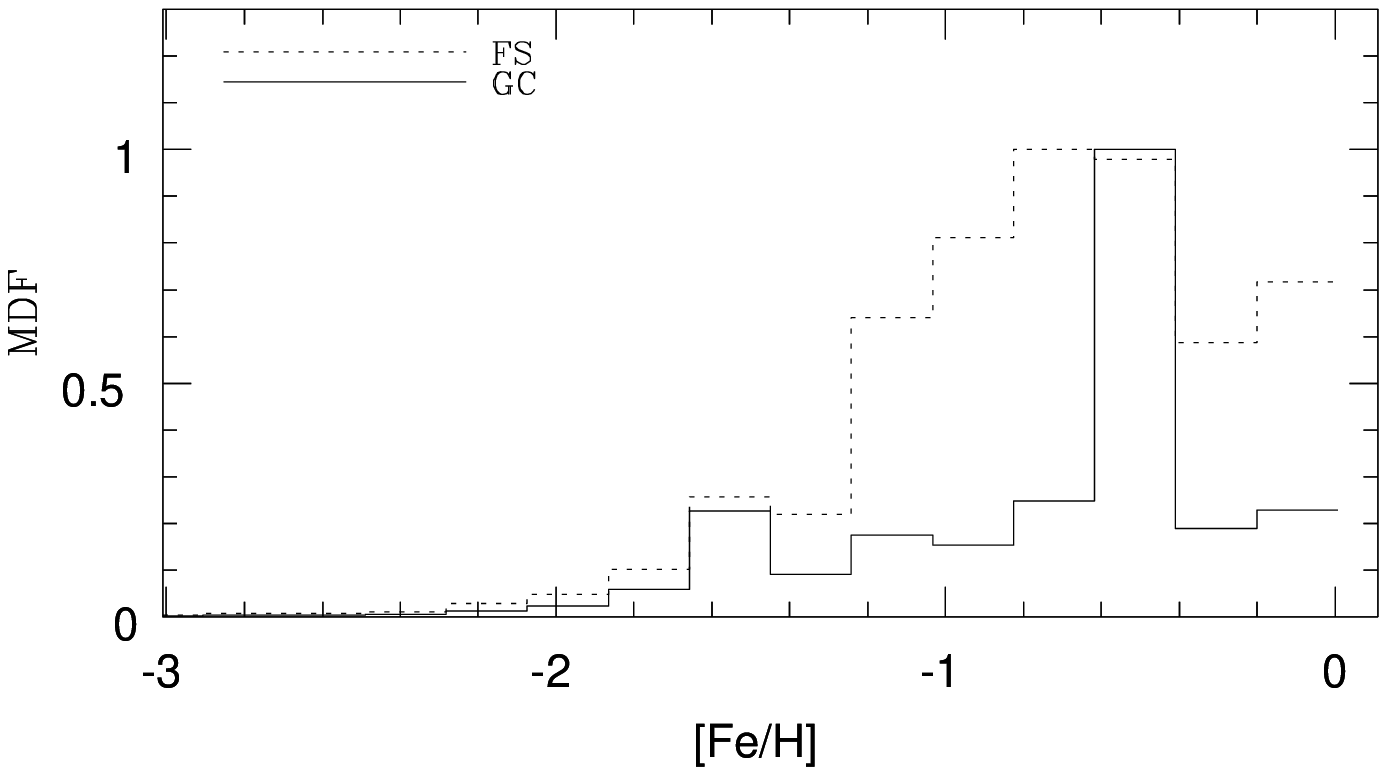,width=8.0cm}
\caption{
The same as Fig. A2  but for the model with $\alpha = 0.02$ 
and $\beta =0.25$.
}
\label{Figure. 28}
\end{figure}

\begin{figure}
\psfig{file=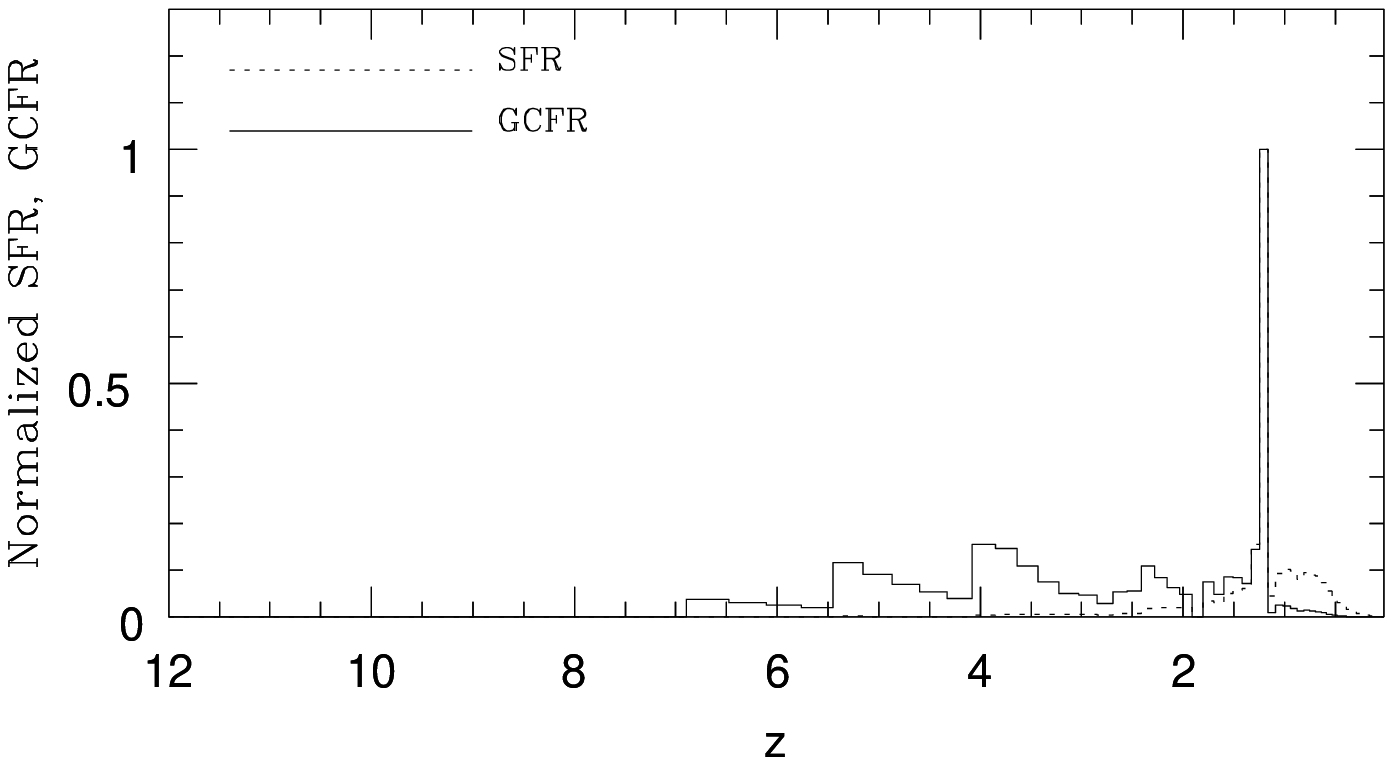,width=8.0cm}
\caption{
The same as Fig. A1  but for galaxy G3.
}
\label{Figure. 29}
\end{figure}

\begin{figure}
\psfig{file=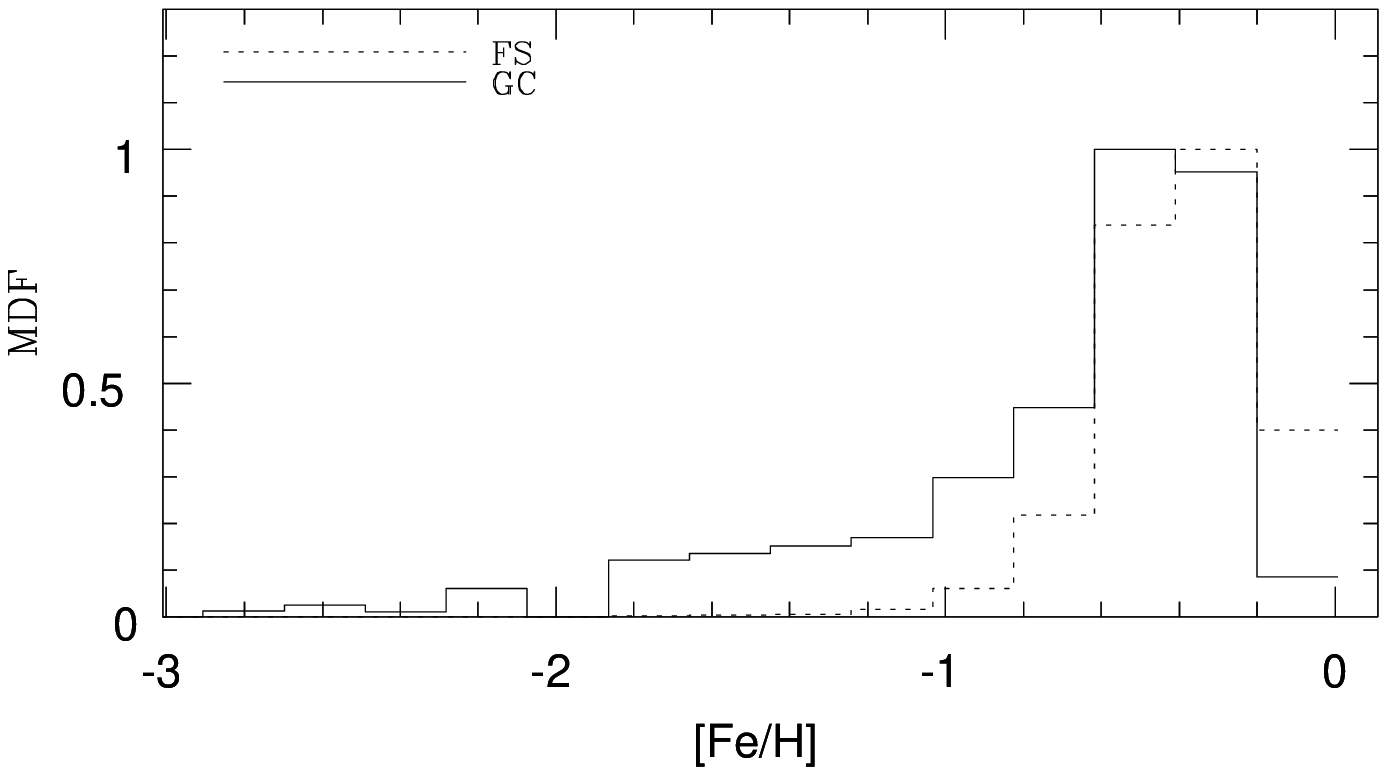,width=8.0cm}
\caption{
The same as Fig. A2  but for galaxy G3.
}
\label{Figure. 30}
\end{figure}

We ran models with different $\alpha$ and $\beta$ in order to more clearly
understand the origin of the bimodal MDFs seen in the present
simulation. Fig. B1 shows the MDF of the galaxy G1 (shown in Figs. A1
and A2) for the model with $\alpha=0.1$ (i.e, five times larger than
that used in the model described in the main text) and $\beta=0.05$.
For this model with very weak dependence of GCFR on $f_{\rm m}$
owing to a larger value of $\alpha$,
strong enhancement in GCFR during major  merging can not happen. 
The MDF, which can be compared with that in Fig. A1, does not
clearly show the metal-rich peak. This means that strong enhancement
in the GCFR during violent merging is important for the appearance 
of bimodality. 

Fig. B2 shows the MDF of G1 for the model with $\alpha=0.02$ and
$\beta=0.25$ (i.e., five times larger than
that used in the model described in the main text).
For this model with very weak dependence of GCFR on $f_{\rm g}$,
strong enhancement in GCFR during gas-rich phases
(thus mostly at high-redshifts) can not happen. 
Although the metal-poor peak can be barely seen,
the bimodal MDF in this model becomes much less significant
than that  in 
the model shown in Fig. A1. This result suggests that
strong enhancement of the  GCFR in gas-rich phases 
of galaxies (thus at high-redshifts) can be also important
for the formation of the bimodality.
Since we assume that galaxies with higher $z_{\rm vir}$ have
higher surface mass densities,
stronger enhancement of GCFR  in more gas-rich galaxies at higher $z$
means high GCFR  in galaxies with higher surface densities of gas. 
These results thus suggest that the origin of the bimodality
can result from stronger enhancement of the GCFR in more violent merging and 
and more gas-rich (and higher surface gas density)
galactic building blocks.

As shown in Fig. 8 and A6, not all of the simulated GCSs
have bimodal MDFs. It is therefore important to
clarify the reasons why these GCSs do not have bimodal MDFs.
By investigating both MDFs and star formation histories  (SFRs)
in galaxies having GCSs without bimodality,
we find that most of GCSs without bimodality
either show only metal-poor  peaks or show only metal-rich ones 
owing to very strong enhancement of the GCFR during single burst epochs 
of GC formation.
Fig. B3 and B4 show the SFH and the MDF, respectively, for the galaxy G3 with
$B/T=0.7$ and $M_{\rm h}=2 \times 10^{11} {\rm M}_{\odot}$.
Owing to the very strong burst of GC formation 
during late major  merging around $z=1$ for this galaxy G3,
too many metal-rich GCs can be formed. This galaxy does not
experience strong enhancement of the GCFR before $z=6$ so that
the metal-poor peak in the MDF is not clear. As  a result of
these, G3 does not show the bimodality in the GCS. 
G3 is the representative case for galaxies which do not show
the bimodality in their GCSs owing to the too many metal-rich
GCs. Other galaxies having GCSs without the bimodality
(which are mostly less-luminous systems)
do not have metal-rich peaks owing to the lack of strong
enhancement of GCFR later in their evolution.
The galaxy labeled as (j) with no strong metal-rich
peak  in Fig. A6 is  a good example of such galaxies.

\end{document}